\documentclass{aa}  
\usepackage{graphicx}
\usepackage{txfonts}
\usepackage{natbib}
\usepackage{lipsum}
\bibpunct{(}{)}{;}{a}{}{,} 
\usepackage{graphicx}
\usepackage{txfonts}
\usepackage[colorlinks=true, citecolor=blue, urlcolor=blue]{hyperref}
\usepackage{placeins}
\usepackage{color, soul}
\begin{document}

   \title{Dissecting the emission from LHAASO J0341+5258: implications for future multi-wavelength observations}


   \author{Agnibha De Sarkar\inst{1}
          \&
          Pratik Majumdar\inst{2}
                 }

   \institute{Astronomy $\&$ Astrophysics group, Raman Research Institute, 
C. V. Raman Avenue, 5th Cross Road, Sadashivanagar, Bengaluru 560080, Karnataka, India
        \and      
             Saha Institute of Nuclear Physics,
A CI of Homi Bhabha National Institute, Kolkata 700064, West Bengal, India
        \\ 
             \email{agnibha@rri.res.in}
             }

   \date{Received YYYY; accepted ZZZZ}

 
  \abstract
   {The Large High Altitude Air Shower Observatory (LHAASO) has detected multiple ultra-high energy (UHE; E$_\gamma \ge$ 100 TeV) gamma-ray sources in the Milky Way Galaxy, which are associated with Galactic ``PeVatrons'' that accelerate particles up to PeV (= 10$^{15}$ eV) energies. Although supernova remnants (SNRs) and pulsar wind nebulae (PWNe), as source classes, are considered the leading candidates, further theoretical and observational efforts are needed to find conclusive proof to confirm the nature of these PeVatrons.}
   {This work aims to provide a phenomenological model to account for the emission observed from the direction of LHAASO J0341+5258, an unidentified UHE gamma-ray source observed by LHAASO. Further, we have also aimed to provide the implications of our model to support future observations in multiple wavelengths.}
   {15 years of Fermi-LAT data was analyzed to find the high energy (HE; 100 MeV $\le$ E$_\gamma$ $\le$ 100 GeV) GeV gamma-ray counterpart of LHAASO J0341+5258, in the 4FGL-DR3 catalog. We have explained the spectrum of the closest 4FGL source, 4FGL J0340.4+5302, by a synchro-curvature emission formalism. Escape-limited hadronic interaction between protons accelerated in an old, now invisible SNR and cold protons inside associated molecular clouds (MCs) and leptonic emission from a putative TeV halo were explored to explain the multi-wavelength (MWL) spectral energy distribution (SED) observed from the LHAASO source region.}
   {The spectrum of 4FGL J0340.4+5302 was explained well by the synchro-curvature emission, which, along with its point-like nature, indicates that it is likely a GeV pulsar. A combined lepto-hadronic emission from SNR+MC and TeV halo scenarios explains the MWL SED of the LHAASO source. We have further found that leptonic emission from an individual TeV halo is also consistent with the observed MWL emission. We have discussed possible observational avenues that can be explored in the near future and predicted the outcome of those observational efforts from the model explored in this paper.}
   {}

   \titlerunning{LHAASO J0341+5258}
   \authorrunning{De Sarkar $\&$ Majumdar}
   \maketitle
%

\section{Introduction}

The nature and emission mechanism of Galactic PeVatrons has become a matter of intense debate after the detection of more than a dozen of UHE gamma-ray sources in the Milky Way Galaxy by LHAASO \citep{cao21a} since it became operational in 2020 April \citep{lhaaso10}. In addition, the successful operations by Tibet-AS$\gamma$ and the High Altitude Water Cherenkov (HAWC) have ushered the era of UHE gamma-ray astronomy \citep{abeysekara20,amenomori19}. Although most of these sources are unidentified, it has been posited that both SNR+MC and PWN/TeV halo systems have the necessary energetics to be the PeVatrons associated with UHE gamma-ray sources. After Crab PWN was confirmed to be a PeVatron \citep{crab21}, the PWN interpretation of PeVatrons was heavily favored. However, recent efforts have suggested that even if a powerful pulsar is present in the vicinity of a UHE gamma-ray source, it is not necessary that the corresponding PWN has to be a PeVatron \citep{desarkar22b}. Furthermore, detailed studies also dictated that SNRs associated with dense MCs are viable candidates for being PeVatrons \citep{desarkar22a,desarkar23,abe23}. Future observational studies by Cherenkov Telescope Array \citep[CTA;][]{CTA19} and the Southern Wide-field Gamma-ray Observatory \citep[SWGO;][]{SWGO19} will be crucial to confirm the nature and emission of PeVatrons.

In this paper, we provide a phenomenological model to explain the MWL emission from the direction of an unidentified UHE gamma-ray source, LHAASO J0341+5258, reported by \cite{cao21c}. This source was detected at the best-fit position of RA = 55.34$^{\circ}$ $\pm$ 0.11$^{\circ}$, and decl. = 52.97$^{\circ}$ $\pm$ 0.07$^{\circ}$, with a significance of 8.2$\sigma$ above 25 TeV. \cite{cao21c} reported that the LHAASO source is spatially extended, where the extension of the source was estimated to be $\sigma_{ext}$ = 0.29$^{\circ}$ $\pm$ 0.06$^{\circ}$, with a TS$_{ext}$ ($\equiv$ 2 log($\mathcal{L}_{ext}$/$\mathcal{L}_{PS}$)) of $\sim$ 13. No apparent energetic pulsar or supernova remnant was found near the LHAASO source. However, from multi-line CO observations ($^{12}$CO and $^{13}$CO) of the region from Milky Way Imaging Scroll Painting (MWISP) project \citep{Su19}, dense MCs were found to be partially overlapped with the LHAASO source. Previously, scenarios including leptonic emission from pulsar halo \citep{cao21c}, hadronic interaction between SNR and MCs \citep{cao21c}, injection of particles from past explosions \citep{kar22} were explored, but none of these models explained the MWL SED entirely. Our simple model aims to provide a feasible MWL emission mechanism to explain the observed MWL SED associated with LHAASO J0341+5258, while accounting for the disappearance of a possible SNR at the present day, as well as the presence of a TeV halo associated with a putative, energetic GeV pulsar within the LHAASO source extent.

In Section \ref{sec:results}, we discuss the results obtained from this work. In Subsection \ref{sec:analysis}, we present the results of Fermi-LAT data analysis of the probable GeV counterpart of the LHAASO source, 4FGL J0340.4+5302. Then in Subsection \ref{sec:synchro}, we provide the basic formalism of the synchro-curvature radiation that has been used to explain the spectrum of the 4FGL source. In Subsections \ref{snr} and \ref{sec:halo}, the models considering the hadronic interaction in the SNR +MC system and the leptonic interaction in the putative TeV halo have been discussed, respectively. Finally, we discuss the results of the study in Section \ref{sec:discussion}, and conclude in Section \ref{sec:conclusion}.


\section{Results}\label{sec:results}

\subsection{Fermi-LAT data analysis}\label{sec:analysis}

15 years (2008 August 4 - 2023 May 1) of PASS 8 Fermi-LAT data in the energy range of 0.1-500 GeV was analyzed using Fermipy\footnote{\url{https://fermipy.readthedocs.io/en/latest/}} version 1.2.0 \citep{wood17}. To avoid contamination from the Earth's albedo gamma rays, the events with a zenith angle greater than 90$^{\circ}$ were excluded from the analysis. The instrument response function, Galactic diffuse emission template (galdiff), and isotropic diffuse emission template (isodiff) used in this analysis were ``P8R3$\_$SOURCE$\_$V3'', ``gll$\_$iem$\_$v07.fits'', and ``iso$\_$P8R3$\_$SOURCE$\_$V3$\_$v1.txt'', respectively. We have used the latest 4FGL catalog, 4FGL-DR3, to study the GeV counterpart of LHAASO J0341+5258 \citep{abdol22}.

A circular Region of Interest (ROI) having a radius of 20$^{\circ}$, with the center coinciding with the centroid of the LHAASO source, was considered to extract the data from the Fermi-LAT website\footnote{\url{https://fermi.gsfc.nasa.gov/ssc/data/access/lat/}}. Within that ROI, a rectangular region of 15$^{\circ}$ $\times$ 15$^{\circ}$, positioned at the centroid of the LHAASO source, was considered. Galdiff, isodiff, and all of the 4FGL sources within that rectangular region were included in the data analysis. The normalization parameters of the 4FGL sources, within 5$^{\circ}$ angular extent of the LHAASO source centroid, including all of the parameters of galdiff and isodiff, were kept free during the data analysis. Previously undetected point sources in the vicinity of the LHAASO source, having a minimum TS value of 25 and a minimum separation of 0.3$^{\circ}$ between any two point sources, were explored using the source-finding algorithm of Fermipy.  However, no plausible point sources relevant to this case were found in the spatial proximity of the LHAASO source. Maximum-likelihood analysis was performed to ascertain the best-fit values of the spatial and spectral parameters of the relevant 4FGL sources, as well as that of galdiff and isodiff. Barring 4FGL J0340.4+5302, which is the probable GeV counterpart of the LHAASO source, the rest of the 4FGL sources, as well as galdiff and isodiff, were considered as background and therefore, subtracted during the analysis. The data analysis procedure discussed above is similar to that followed in \cite{desarkar22c}.  

\cite{cao21c} has analyzed 4FGL-DR2 data, and found the same GeV counterpart 4FGL J0340.4+5302, within the extension of LHAASO J0341+5258. We have rechecked the properties of the 4FGL source with updated 4FGL-DR3 data to ascertain the localization, extension, and spectrum of the source. The 4FGL source was located at RA = 55.135$^{\circ}$ $\pm$ 0.013$^{\circ}$ and decl. = 53.083$^{\circ}$ $\pm$ 0.011$^{\circ}$ with a significance of 64.61$\sigma$, 0.154$^{\circ}$ away from the centoid of the LHAASO source. Similar to \cite{cao21c}, the spectrum of the 4FGL source was found to be significantly curved (TS$_{curve}$ $\equiv$ 2 log($\mathcal{L}_{LP}$/$\mathcal{L}_{PL}$) $\sim$ 325.74) and well-fitted by a log-Parabola spectrum, i.e., dN/dE $\propto$ (E/E$_b$)$^{-\alpha_{LP}-\beta_{LP}\:log(E/E_b)}$, with best-fit spectral parameters, $\alpha_{LP}$ = 3.106 $\pm$ 0.047, $\beta_{LP}$ = 0.483 $\pm$ 0.033, E$_b$ = 0.541 GeV, and the corresponding energy flux is $\sim$ 5.447 $\times$ 10$^{-11}$ erg cm$^{-2}$ s$^{-1}$ in the energy range of 0.1-500 GeV. The source extension was checked with a \texttt{RadialDisk} model. The 95$\%$ confidence level upper limit of the extension of the 4FGL source was found to be, $\sigma_{disk}$ $\leq$ 0.29$^{\circ}$, with TS$_{ext}$ of $\sim$ 15.08 (3.88$\sigma$), indicating that the 4FGL source is a point-like source. Due to point-like extension and curved spectral signature associated with the 4FGL source, we posit that 4FGL J0340.4+5302 is possibly a pulsar emitting in GeV gamma-ray range. This conclusion was also echoed in the work done by \cite{cao21c}. 

\subsection{Synchro-curvature emission from putative pulsar}\label{sec:synchro}

To test the GeV pulsar interpretation of the 4FGL source, we explore the synchro-curvature emission formalism, which has been previously used to explain GeV gamma-ray emission from pulsars \citep{cheng96,kelner15}. The GeV gamma-ray emission from energetic pulsars has been conventionally explained by two general mechanisms: (a) Curvature emission, where the radiation is produced by relativistic electron-positron pairs streaming along the curved magnetic field lines with a radius of curvature, and (b) Synchrotron emission, where the radiation is produced by the same pairs gyrating around a straight magnetic field line. Although both of these emission mechanisms explain the GeV gamma-ray emission from pulsars well, in a realistic scenario, it can be clearly understood that the relativistic charged particles streaming along the curved magnetic field lines, must also spiral around them. Consequently, rather than proceeding in either the curvature or the synchrotron radiation modes, an intermediate emission scenario termed the synchro-curvature radiation, should be considered the general radiation mechanism responsible for gamma-ray observed from GeV pulsars (for further details, see \cite{cheng96}, \cite{vigano15}). Hence, in this work, we try to explain the spectrum of the 4FGL source with the synchro-curvature process, assumed to happen in the outer gap of the pulsar magnetosphere. In this section, we outline the governing equations relevant to the synchro-curvature radiation formalism. For a detailed discussion on the topic, please refer to \cite{cheng96,vigano15,vigano15b,vigano15c}.

The particles, spiraling around a curved magnetic field with a radius of curvature r$_c$ and magnetic field B, emit photons with characteristic energy,

\begin{equation}\label{eq1}
    E_c(\Gamma, r_c, r_{gyr}, \alpha) = \frac{3}{2} \hbar c Q_2 \Gamma^3
\end{equation}

where, $\Gamma$ is the relativistic Lorentz factor, $\alpha$ is the pitch angle (angle between $\Vec{B}$ and $\Vec{v}$), $\hbar$ ($\approx$ 1.0546 $\times$ 10$^{-27}$ cm$^2$ g s$^{-2}$ K$^{-1}$) is the reduced Planck's constant. Gyro-radius (or Larmor radius) r$_{gyr}$ and the factor Q$_2$ are given by,

\begin{equation}\label{eq2}
    r_{gyr} = \frac{m_e c^2 \Gamma sin\:\alpha}{e B}
\end{equation}

\begin{equation}\label{eq3}
    Q_2^2 = \frac{cos^4 \alpha}{r_c^2}\left[1 + 3\xi + \xi^2 + \frac{r_{gyr}}{r_c} \right]
\end{equation}

where m$_e$ is the electron rest mass, and c is the velocity of light. The synchro-curvature parameter $\xi$ is given by,

\begin{equation}\label{eq4}
    \xi = \frac{r_c}{r_{gyr}} \frac{sin^2 \alpha}{cos^2 \alpha}
\end{equation}

The power radiated by a single particle per unit energy at a given position is given by,

\begin{equation}\label{eq5}
    \frac{d P_{sc}}{d E} = \frac{\sqrt{3} e^2 \Gamma y}{4 \pi \hbar r_{eff}}\left[(1+z)F(y) - (1-z)K_{2/3}(y) \right]
\end{equation}

where, 

\begin{equation}\label{eq6}
    y(E, \Gamma, r_c, r_{gyr}, \alpha) \equiv \frac{E}{E_c}
\end{equation}

\begin{equation}\label{eq7}
    z = (Q_2 r_{eff})^{-2}
\end{equation}

\begin{equation}\label{eq8}
    F(y) = \int^{\infty}_{y} K_{5/3}(y')\:dy'
\end{equation}

where E is the photon energy, K$_n$ are the modified Bessel functions of the second kind of index n, and the effective radius is given by,

\begin{equation}\label{eq9}
    r_{eff} = \frac{r_c}{cos^2 \alpha}\left(1 + \xi + \frac{r_{gyr}}{r_c}\right)^{-1}
\end{equation}

By integrating Equation \ref{eq5} in energy, we get the total synchro-curvature power radiated by a single particle,

\begin{equation}\label{eq10}
    P_{sc} = \frac{2 e^2 \Gamma^4 c}{3 r_c^2} g_r
\end{equation}

where synchro-curvature correction factor g$_r$ is given by,

\begin{equation}\label{eq11}
    g_r = \frac{r_c^2}{r_{eff}^2} \frac{\left[1 + 7(r_{eff} Q_2)^{-2}\right]}{8 (Q_2 r_{eff})^{-1}}
\end{equation}

We have further obtained the details regarding the trajectories of the charged particles by numerically solving their equations of motion,

\begin{equation}\label{eq12}
    \frac{d\Vec{p}}{dt} = eE_{||}\hat{b} - \frac{P_{sc}}{v}\hat{p}
\end{equation}

In this equation, the relativistic momentum (with the velocity assumed to be constant at v=c) of the charged particles, $\Vec{p}$ (= $\sqrt{\Gamma^2 - 1}$mc$\hat{p}$ = $\Gamma$mv$\hat{p}$), is directed towards $\hat{p}$, and the constant accelerating electric field, $E_{||}$, is directed towards $\hat{b}$, i.e., tangential to the curved magnetic field lines. Breaking down the equations of motion into parallel (p$_{||}$ = p cos $\alpha$) and perpendicular (p$_{\perp}$ = p sin $\alpha$) components, we get,

\begin{equation}\label{eq13}
    \frac{d(p\:sin\:\alpha)}{dt} = - \frac{P_{sc}\:sin\:\alpha}{v}
\end{equation}

\begin{equation}\label{eq14}
    \frac{d(p\:cos\:\alpha)}{dt} = eE_{||} - \frac{P_{sc}\:cos\:\alpha}{v}
\end{equation}

Equations \ref{eq13} and \ref{eq14} are numerically solved to determine the evolution of the Lorentz factor $\Gamma$, sin $\alpha$, and synchro-curvature parameter $\xi$ along the trajectory of motion.

Similar to \cite{vigano15}, we calculate the average synchro-curvature radiation spectrum throughout the trajectory using the equation,

\begin{equation}\label{eq15}
    \frac{d P_{tot}}{dE} = \int^{x_{max}}_0 \frac{d P_{sc}}{dE} \frac{d N}{dx} dx
\end{equation}

where the integration limits have been chosen to be the distance depicting the injection point of the particles (x=0), and the maximum distance up to which the spectrum can be emitted (x=x$_{max}$). Furthermore, the effective weighted particle distribution function, which takes into account the depletion of the number of emitting particles directed toward the observer at a distance x from their injection point, is given by \citep{vigano15},

\begin{equation}\label{eq16a}
    \frac{dN}{dx} = \frac{N_0\:e^{-x/x_0}}{x_0(1 - e^{-x_{max}/x_0})}
\end{equation}

Here, N$_0$, the normalization of the effective particle distribution, is such that $\int^{x_{max}}_0$(dN/dx)dx = N$_0$, and x$_0$ is the length scale of the same.

The model discussed above is based on the dynamics of relativistic lepton pairs that move along curved magnetic field lines in an acceleration region of the pulsar magnetosphere. The calculation has been done considering three free parameters:

\begin{enumerate}
    
    \item The electric field parallel to the magnetic field, E$_{||}$ (V m$^{-1}$), which is assumed to be constant throughout the acceleration region. This parameter has been varied within the range log(E$_{||}$ (V m$^{-1}$)) = 6.5 -- 9.5 \citep{vigano15b}. The accelerating electric field explains the energy peak of the synchro-curvature spectrum.
 
    \item The length scale, x$_0$/r$_c$, which depicts the spatial extent of the emitting region for injected particles. The parameter has been varied within the range x$_0$/r$_c$ = 0.001 -- 1 \citep{vigano15b}. The variation of this parameter determines the low-energy slope of the spectrum.  

    \item The overall normalization parameter, N$_0$, which depicts the total number of charged particles in the acceleration region, whose radiation is directed toward the observer. The overall normalization N$_0$ has been varied to explain the spectrum of the 4FGL source. The parameter has been varied within the range of N$_0$ = 10$^{26}$ -- 10$^{34}$ \textit{particles} \citep{vigano15b}. 

\end{enumerate}

The rest of the parameters are considered to be fixed following \cite{vigano15}, i.e., the magnetic field B = 10$^6$ G, radius of curvature r$_c$ = 10$^8$ cm, maximum distance of emitting region x$_{max}$ = r$_c$ = 10$^8$ cm. Two coupled ordinary differential equations, equations \ref{eq13} and \ref{eq14}, are numerically solved simultaneously to evaluate the evolution of Lorentz factor $\Gamma$, pitch angle in terms of sin $\alpha$, and the synchro-curvature parameter $\xi$. To solve these equations, the initial values for Lorentz factor and pitch angle have been typically set to be $\Gamma_{in}$ = 10$^3$ and $\alpha_{in}$ = 45$^{\circ}$ \citep{vigano15}. Note that although the magnetic field can be ideally parameterized as a function of the timing properties and the magnetic gradient \citep{vigano15b,vigano15c}, due to a lack of knowledge regarding those parameters in this case, we consider the magnetic field to be constant at a value consistent with that explored in \cite{vigano15}.

\begin{figure*}[htp]
\centering
\includegraphics[width=.47\textwidth]{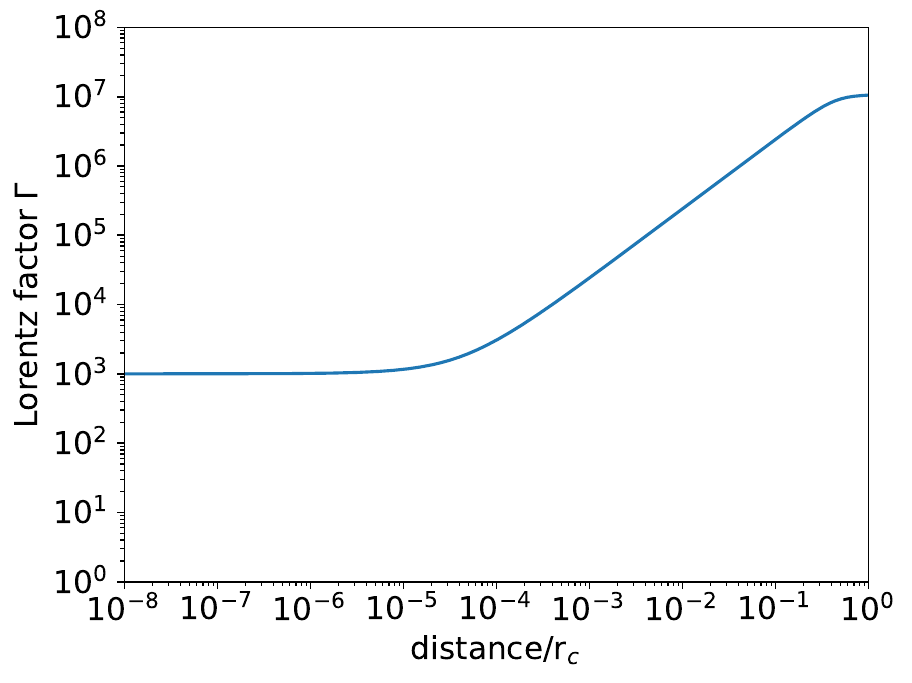}(a)
\includegraphics[width=.47\textwidth]{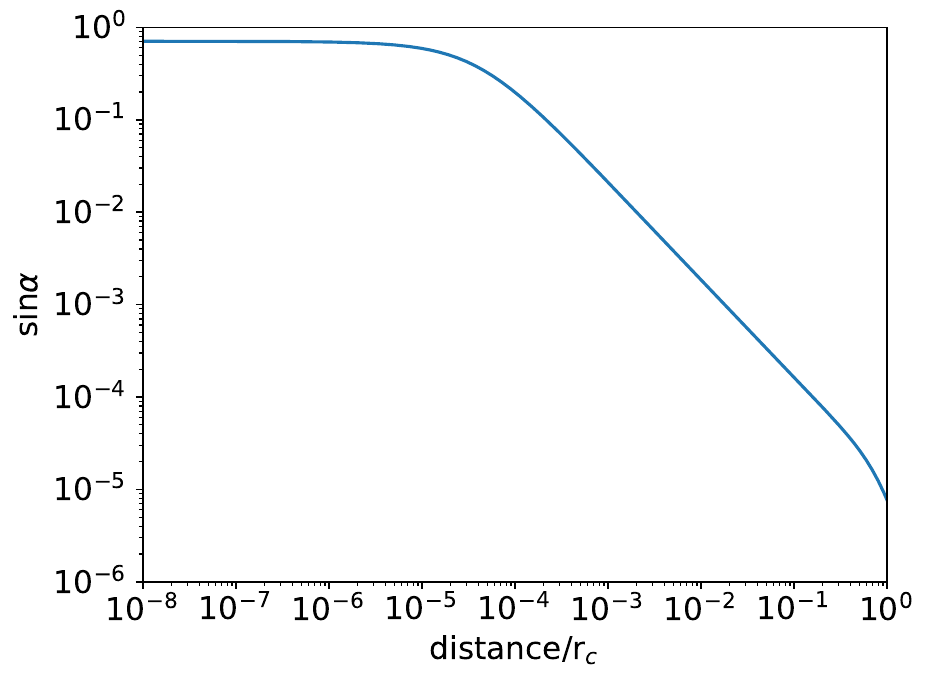}(b)
\includegraphics[width=.47\textwidth]{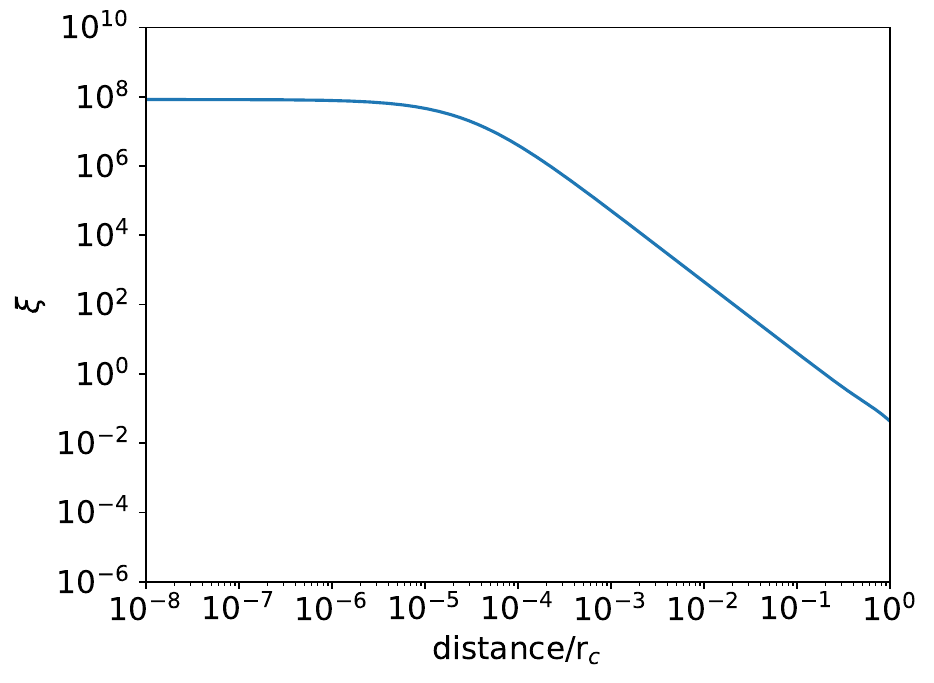}(c)
\includegraphics[width=.47\textwidth]{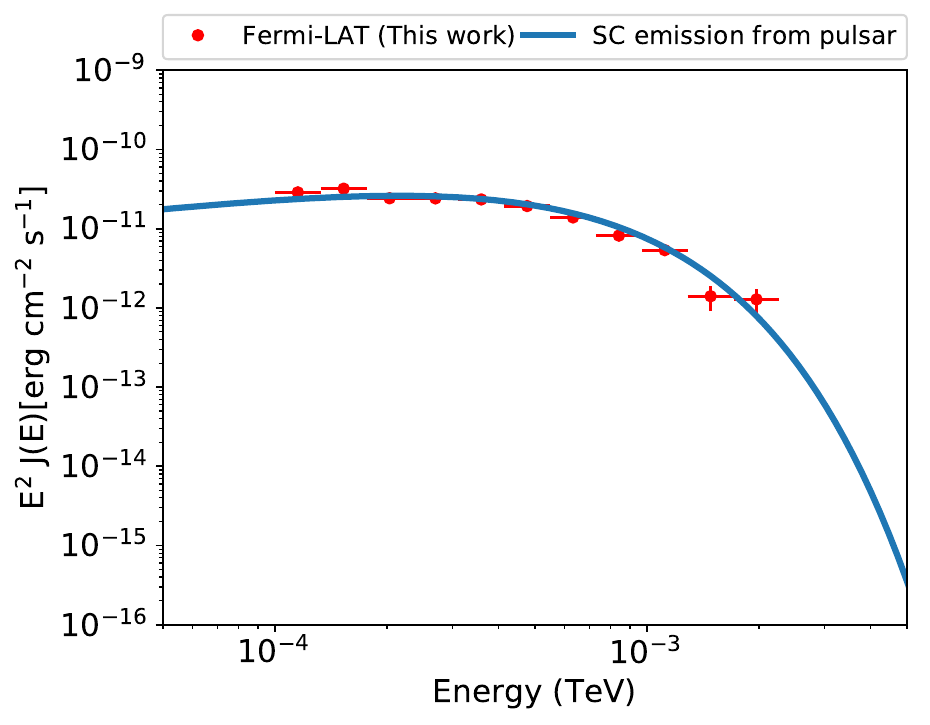}(d)
\caption{In the Figure, the plots corresponding to the outputs of the synchro-curvature model are given. The evolution of (a) Lorentz factor $\Gamma$, (b) pitch angle $\alpha$, and (c) synchro-curvature parameter $\xi$ are given. In panel (d), the model spectrum is plotted against the SED data points obtained from Fermi-LAT analysis of 4FGL J0340.4+5302.}
\label{fig1}    
\end{figure*}

In Figure \ref{fig1}, the evolution of Lorentz factor $\Gamma$ (panel (a)), pitch angle $\alpha$ (panel (b)), synchro-curvature parameter $\xi$ (panel (c)), and model spectrum against the SED data points of the 4FGL source (panel (d)) are plotted. The values of the free parameters considered in this model to explain the SED of 4FGL J0340.4+5302 are log(E$_{||}$ (V m$^{-1}$)) = 7.113, x$_0$/r$_c$ = 0.15, and N$_0$ = 1.3 $\times$ 10$^{31}$ \textit{particles}, where the distance to the pulsar was assumed to be 1 kpc \citep{cao21c}. From panel (d) of Figure \ref{fig1}, it can be seen that the synchro-curvature emission model explains the SED of the 4FGL source quite well, which, in turn, indicates that 4FGL J0340.4+5302 indeed shows typical spectral features of a GeV pulsar. Detection of pulsed emission from this source in radio and gamma rays would confirm its nature in the future.

\subsection{Emission from SNR+MC association}\label{snr}

In this section, we discuss the full model and the relevant parameters of the hadronic interaction model, in which gamma rays are produced from inelastic p-p interaction between protons accelerated in the shock front of an old, now invisible, shell-type SNR and the cold protons residing in the MCs surrounding the SNR. We have used open source code \texttt{GAMERA}\footnote{\url{http://libgamera.github.io/GAMERA/docs/main_page.html}} \citep{hahn} to calculate gamma-ray SED from the hadronic p-p interaction. For a detailed discussion on the formalism, please refer to \cite{desarkar22a}, \cite{desarkar23}, \cite{fujita09}, \cite{ohira10}, \cite{makino19}.

The model assumes that a supernova (SN) explosion had occurred inside a tenuous, spherical cavity, surrounded by dense MCs. After the explosion, following the initial free expansion phase, the SNR enters the adiabatic Sedov-Taylor phase, during which the time evolution of the shock velocity and shock radius is given by the relations \citep{desarkar22a, fujita09},   

\begin{equation}
\label{eq16}
\begin{split}
v_{sh}(t) = 
\begin{cases}
v_i & (t < t_{Sedov})\\
v_i(t/t_{Sedov})^{-3/5} & (t_{Sedov} < t)\\
\end{cases}              
\end{split}
\end{equation}

and, 

\begin{equation}
\label{eq17}
\begin{split}
R_{sh}(t) \propto 
\begin{cases}
(t/t_{Sedov}) & (t < t_{Sedov})\\
(t/t_{Sedov})^{2/5} & (t_{Sedov} < t)\\
\end{cases}              
\end{split}
\end{equation}

where the initial shock velocity v$_i$ = 10$^9$ cm s$^{-1}$, SNR age and radius at the onset of the Sedov phase, t$_{Sedov}$ $\approx$ 210 years and R$_{Sedov}$ $\approx$ 2.1 pc were assumed. The cosmic ray (CR) protons are accelerated through Diffusive Shock Acceleration (DSA) at the shock front. We adopt an escape-limited scenario of proton acceleration \citep{ohira10}, where these accelerated protons need to escape a geometrical confinement region around the shock front, produced by strong magnetic turbulence, to participate in gamma-ray production after the shock front collides with the surrounding MCs. The distance of the outer boundary of this confinement region (escape boundary) from the center of the cavity, i.e., the escape radius, is given by,

\begin{equation}
\label{eq18}
R_{esc}(t) = (1 + \kappa) R_{sh}(t),
\end{equation}

where $\kappa$ ($\approx$ 0.04) is defined by the relation l$_{esc}$ = $\kappa$R$_{sh}$, where l$_{esc}$ is the radial distance of the escape boundary from the shock front \citep{ohira10, makino19}.

It has been assumed that the acceleration of protons stops at the time of collision t = t$_{coll}$, i.e., when the escape radius is equal to the distance of the MC surface from the cavity center (i.e., R$_{esc}$ (t$_{coll}$) $\approx$ R$_{sh}$ (t$_{coll}$) = R$_{MC}$) \citep{fujita09}. So only the protons, which have been accelerated before the collision and possess sufficient energy to escape the escape boundary, will take part in producing UHE gamma rays. This threshold energy of proton escape can be given by the phenomenological relation \citep{makino19, ohira12},

\begin{equation}
\label{eq19}
E_{esc} = E^{max}_{SNR} \left(\frac{R_{sh}}{R_{Sedov}}\right)^{-\alpha_{SNR}},   
\end{equation}

where $\alpha_{SNR}$ signifies the evolution of the escape energy during the Sedov phase \citep{makino19}. Note that in this case, it has been assumed that the protons get accelerated up to a maximum energy of E$^{max}_{SNR}$ $\approx$ 10$^{15.5}$ eV (knee energy) at the onset of the Sedov phase \citep{gabici09}. We consider $\alpha_{SNR}$ as a free parameter, and E$^{min}_{SNR}$ = E$_{esc}$, where E$^{min}_{SNR}$ is the minimum energy of the escaped proton population. The spectrum of the escaped proton population is given by \citep{ohira10},

\begin{equation}
\label{eq20}
    N_{esc}(E_p) \propto E_p^{-[s + (\beta/\alpha_{SNR})]} \propto E_p^{-p_{SNR}},
\end{equation}

where $\beta$ = 3(3–s)/2 \citep{makino19}, assuming the thermal leakage model of CR injection \citep{ohira10}. For s = 2, as is expected from DSA, we find $\beta$ = 1.5. Note that the minimum energy (equation \ref{eq19}), and the spectral shape (equation \ref{eq20}) of the escaped proton population, as well as the gamma-ray production from the hadronic p-p interaction \citep{kafe}, are all estimated at the collision time t = t$_{coll}$. In this particular work, the value of the free parameter $\alpha_{SNR}$ was phenomenologically varied and was chosen to be $\alpha_{SNR}$ = 1.5. Considering the chosen value of $\alpha_{SNR}$, our model indicates that the expanding SNR shock collided with the surrounding dense MCs at an age of t$_{coll}$ $\sim$ 6.1 $\times$ 10$^3$ years. At time t = t$_{coll}$, the radius and the velocity of the SNR shock front were found to be R$_{sh}$ (t$_{coll}$) $\sim$ 20.27 pc (which is also equal to R$_{MC}$ at the time of collision), and v$_{sh}$ (t$_{coll}$) $\sim$ 1.3 $\times$ 10$^8$ cm s$^{-1}$, respectively. Following the collision, the escaped proton population accelerated until the collision epoch, seeps inside the MC medium to produce gamma rays through hadronic p-p interaction. The minimum energy of this escaped proton population is found to be E$^{min}_{SNR}$ $\sim$ 100 TeV, calculated using equation \ref{eq19} for the choice of the parameter $\alpha_{SNR}$, whereas, as discussed above, the maximum energy is given by E$^{max}_{SNR}$ $\sim$ 3.1 $\times$ 10$^3$ TeV. Furthermore, using values of s, $\beta$, and $\alpha_{SNR}$, the spectral index of the escaped proton population was calculated to be p$_{SNR}$ = 3.0, and the corresponding spectral shape was given by equation \ref{eq20}. The total energy budget of this escaped proton population required to explain the gamma-ray SED was found to be W$_{SNR}$ $\sim$ 1.7 $\times$ 10$^{46}$ erg, where the number density inside the MC medium and the SNR+MC source distance was assumed to be n$_{MC}$ $\sim$ 50 cm$^{-3}$ and d = 1 kpc respectively, following \cite{cao21c}. 

\begin{figure*}[htp]
\includegraphics[width=.49\textwidth]{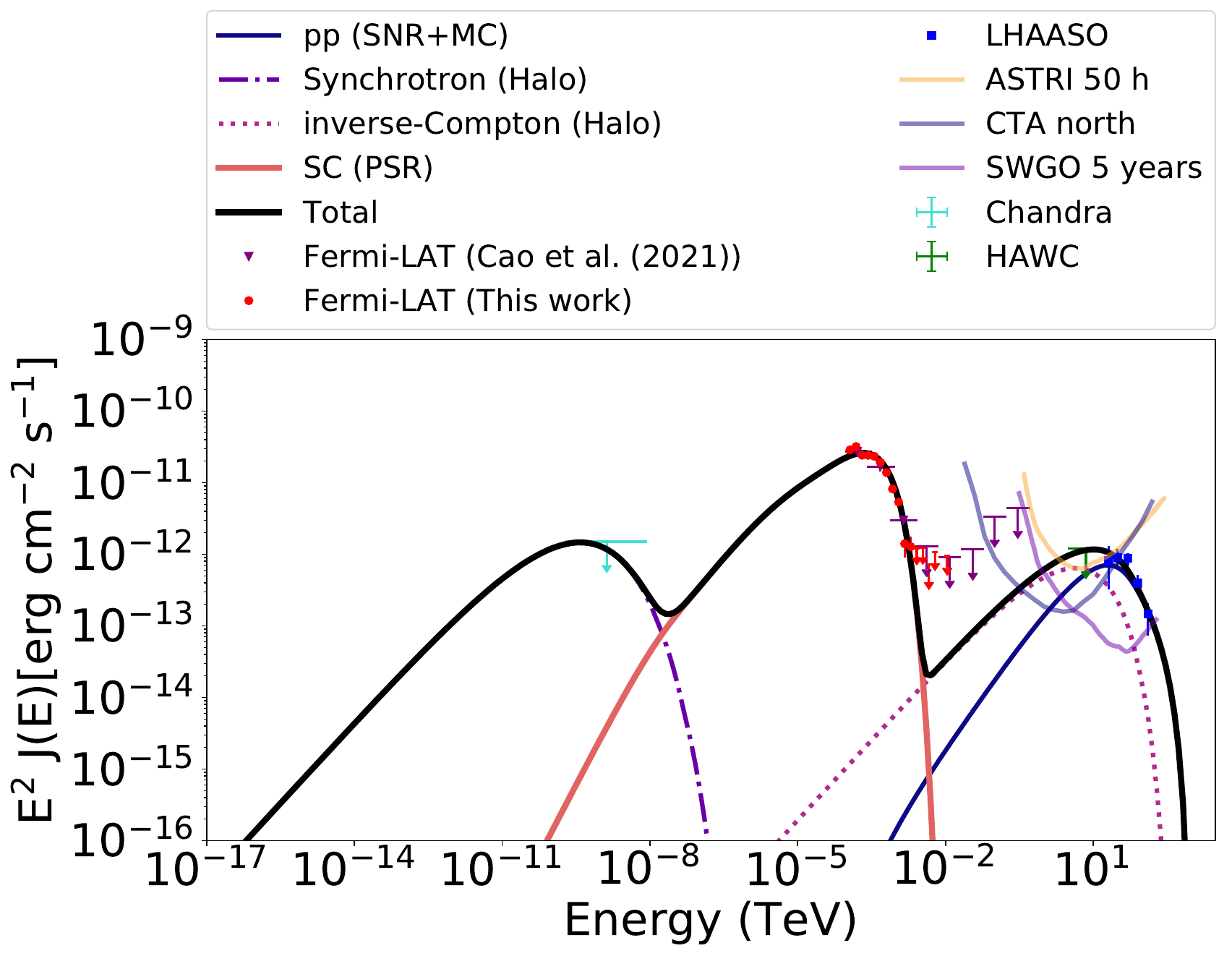}(a)
\includegraphics[width=.49\textwidth]{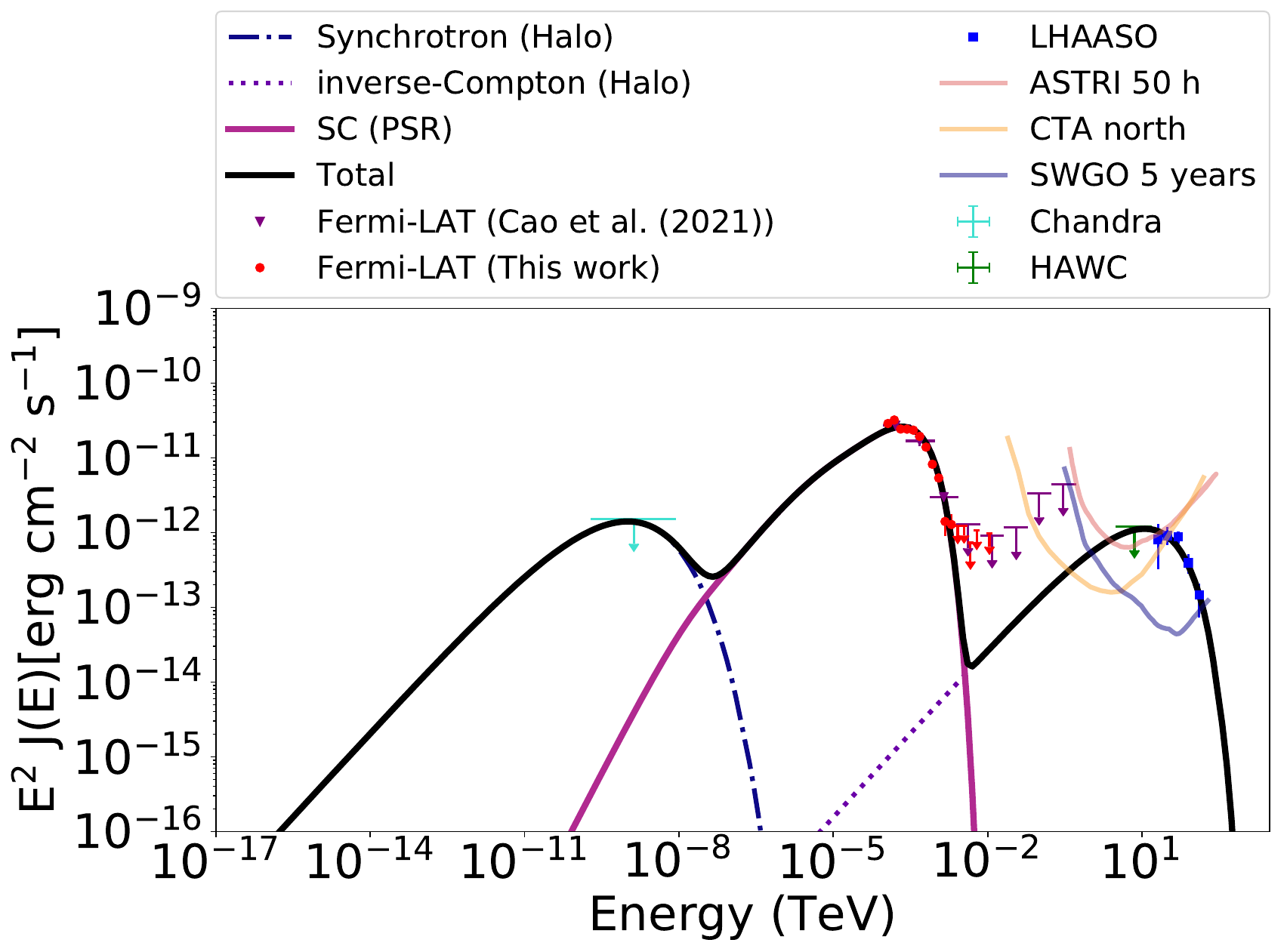}(b)
\caption{In the Figure, the plots containing the MWL data points, along with the MWL spectra obtained from the two models discussed in this paper, are provided. The (a) Lepto-hadronic model spectrum from combined SNR+MC and TeV halo scenarios, and (b) Leptonic model spectrum from a single TeV halo scenario are plotted against the MWL SED of LHAASO J0341+5258.}
\label{fig2}    
\end{figure*}

\begin{figure}[htp]
\centering
\includegraphics[width=\columnwidth]{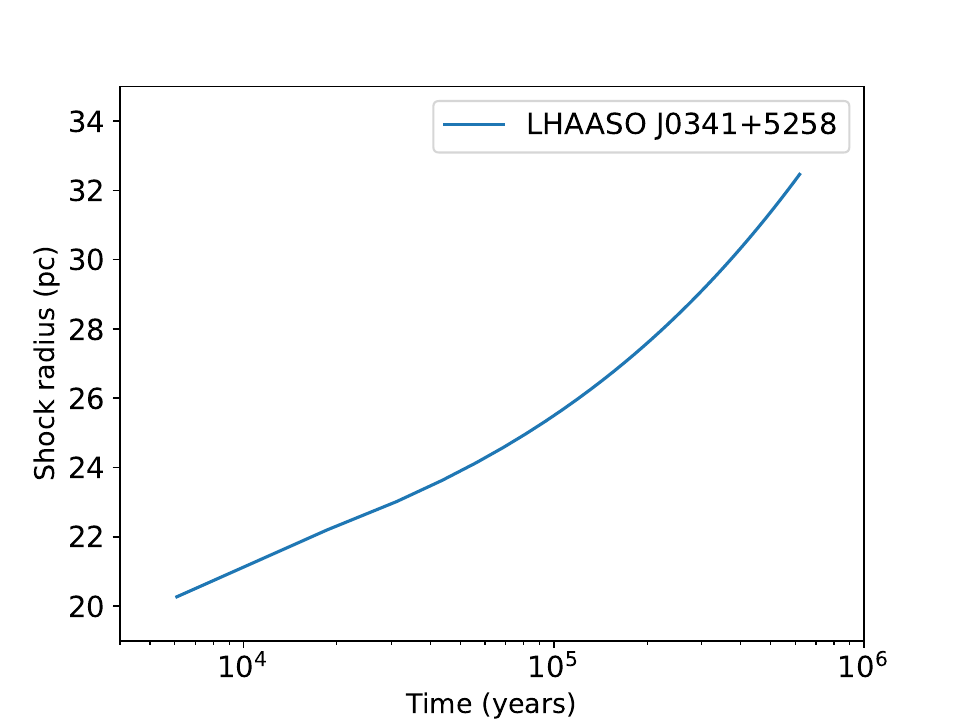}
\caption{The time evolution of the shocked shell associated with the old SNR inside the surrounding MCs is plotted.}
\label{fig3}
\end{figure}

\begin{figure}[htp]
\centering
\hspace{-1.4cm}\includegraphics[width=\columnwidth]{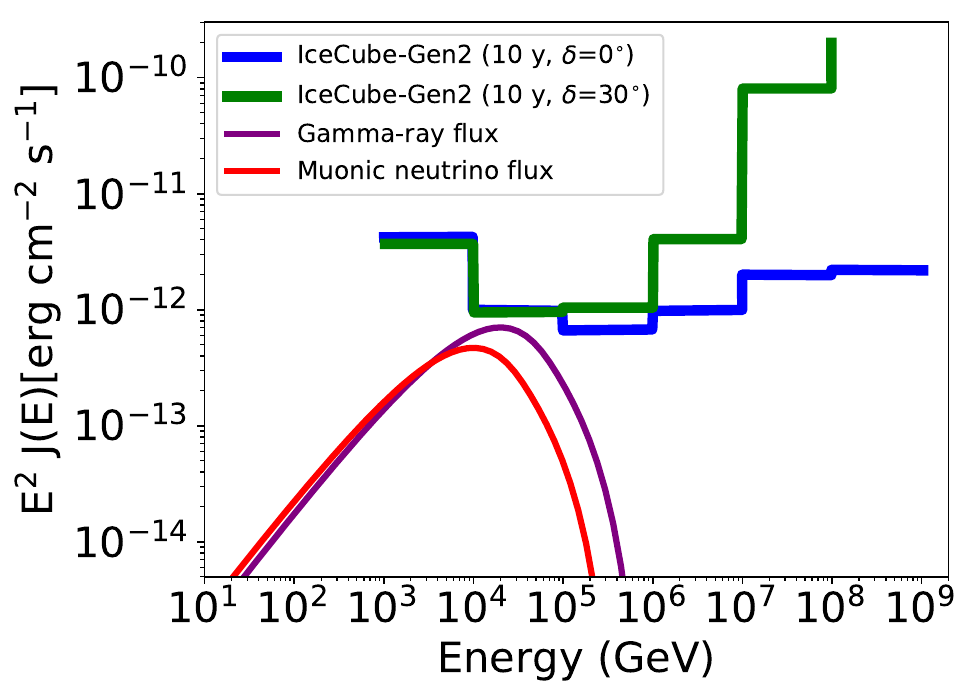}
\caption{The expected neutrino flux (scaled) is plotted against the IceCube-Gen2 sensitivity for two declinations.}
\label{fig4}
\end{figure}

At t = t$_{coll}$, the shock can be assumed as a shell with a radius of R$_{sh}$(t$_{coll}$) (= R$_{MC}$), centered at the cavity. At t $>$ t$_{coll}$, the shock enters the momentum-conserving, snow-plow phase, and continues to expand inside the MC medium. If the radius of the shell inside the MC medium is R$_{shell}$, then its time evolution inside the MCs can be estimated by solving the momentum conservation equation \citep{fujita09, desarkar22a},

\begin{equation}
\label{eq21}
\begin{split}
\frac{4\pi}{3}&\left[n_{MC} ( R_{shell}(t)^3 -  R_{sh}(t_{coll})^3) + n_{cav}R_{sh}(t_{coll})^3\right]\Dot{R}_{shell}(t)\\
& = \frac{4\pi}{3}n_{cav}R_{sh}(t_{coll})^3 v_{sh}(t_{coll}), 
\end{split}
\end{equation}

with R$_{shell}$ = R$_{MC}$ at t = t$_{coll}$, and n$_{cav}$ ($\approx$ 1 cm$^{-3}$) is the number density inside the cavity. Note that the velocity of the shocked shell inside the MC medium continues to decrease as it continues to expand with time. As a result, if the SNR shocked shell at the current epoch is old enough, its velocity inside the MCs will definitely be comparatively smaller than the internal gas velocity of the MCs. Consequently, the shocked shell inside the MCs will not be detectable as the remains of the shell will become invisible. We use this fact to explain the non-detection of the possible old SNR and to posit the probable current age of the SNR as well. This approach was used to explain the non-detection of the SNR shell in the case of LHAASO J2108+5157 \citep{desarkar23}. We calculate the time evolution of SNR shocked shell inside the associated MCs using equation \ref{eq21}, and find that the SNR, with a final radius of R$_{sh}$ (t$_{age}$) $\sim$ 32.4 pc, has to be t$_{age}$ $\sim$ 6.2 $\times$ 10$^5$ years old, for the shock velocity (v$_{sh}$ (t$_{age}$) $\sim$ 8 $\times$ 10$^5$ cm s$^{-1}$) to be lower than the internal gas velocity of MCs ($\sim $10$^6$ cm s$^{-1}$; \cite{cao21c}), and the SNR shell to disappear. The time evolution of the shocked shell is shown in Figure \ref{fig3}. Please note that we do not consider the total gamma-ray flux produced from the escaped protons, when the shock front is within the MC medium, even if the SNR is still in the Sedov phase. The acceleration and escape of protons will depend on the evolution of the confinement region inside the turbulent MC medium, which is poorly understood. Consequently, we have avoided this contribution altogether not to complicate our model, as this contribution is expected to be negligible anyway.  Moreover, due to a small shock velocity, the full ionization of the pre-shock gas does not occur, making the particle acceleration ineffective when the SNR enters the radiative phase. As a result, the corresponding gamma-ray contribution during the radiative phase of the SNR continues to remain insignificant (see \cite{desarkar23} and references therein). We further note that proton diffusion inside the MC medium has been neglected in this model. The average diffusion coefficient inside the dense, strongly turbulent MC medium ($\approx$ 10$^{25}$ - 10$^{26}$ cm$^2$s$^{-1}$ \citep{gabici09}) is significantly smaller than that measured in the interstellar medium ($\approx$ 10$^{28}$ - 10$^{29}$ cm$^2$s$^{-1}$ \citep{desarkar21}). The details regarding the suppressed diffusion inside the MCs are uncertain \citep{dogiel15,xu16}, so we exclude this aspect to avoid introducing complications in the simple model discussed in this paper. A similar assumption was also considered in the case of LHAASO J1908+0621 \citep{desarkar22a} and for LHAASO J2108+5157 \citep{desarkar23}. 

Note that neutrino emission is a smoking gun evidence for hadronic interaction in any astrophysical source. So, to confirm the presence of a hadronic emission mechanism in this particular source, we compared the total neutrino flux expected from hadronic interaction to the sensitivity of the next-generation IceCube-Gen2 neutrino observatory \citep{aartsen21}. We found that the neutrino flux is not significant enough to be detected by IceCube-Gen2. We have plotted the scaled neutrino flux, along with IceCube-Gen2 sensitivity, in Figure \ref{fig4}.

\subsection{Emission from TeV halo}\label{sec:halo}

As an energetic pulsar spins down, a wind nebula is created due to the conversion of rotational energy to wind energy, known as pulsar wind nebula (PWN) \citep{gaensler06}. Electron-positron pairs, that got accelerated to ultra-relativistic energies at the termination shock of the wind, produce MWL emission due to interaction with the ambient magnetic field, matter, and radiation fields. As a result, throughout the years, multiple PWNe have been detected, especially in radio, X-ray, and gamma-ray energy ranges \citep{gaensler06}, and PWNe are considered to be one of the leading candidates for being Galactic PeVatrons \citep{emma22}. The size of the PWNe can be of the order of 0.1 - 10 pc, and the associated nebular magnetic field can be estimated to be of the order of 10 - 1000 $\mu$G. PWN is a dynamic source class, which goes through multiple stages of evolution \citep{giacinti20}. In the first stage (t $<$ 10 kyr), PWNe can be considered as a spherically symmetric system, in which high energy leptons are confined due to a large magnetic field, and TeV gamma rays are emitted by these leptons. The forward shock of the host SNR expands in the surrounding ISM, whereas the newly formed reverse shock starts to contract, but does not yet reach the PWN. In the second stage (t = 10 - 100 kyr), the PWN morphology becomes highly irregular, as, at this stage, the reverse shock has hit the PWN, thus disrupting it. At this stage, the high-energy leptons escape and propagate inside the surrounding SNR, but not yet in the surrounding interstellar medium (ISM). In the final stage (t $>$ 100 kyr), the nebula completely disrupts and the host SNR fades away. The high-energy leptons thus escape in the surrounding ISM, and then slowly diffuse in the strongly turbulent interstellar magnetic field and emit TeV gamma rays in a volume that is much larger than that of the initial PWN.

This extended source class, associated with energetic pulsars, emitting very high energy (VHE; 100 GeV $\le$ E$_\gamma$ $\le$ 100 TeV) gamma rays, known as TeV halo, has recently been established, which shines bright in TeV energies and has a hard spectrum (having an electron injection spectral index between $\sim$ 1.5 to 2.2 \citep{sudoh19}). TeV halos were first detected by the MILAGRO and HAWC observations of Geminga and PSR B0656+14, where extended TeV gamma-ray emission was discovered surrounding these pulsars, from the surface brightness distributions \citep{milagro09, abey17a, abey17b}. TeV halos are characterized by a slow diffusion region (e.g., D(E$_e$) = 4.5 $\times$ 10$^{27}$ (E$_e$/100 TeV)$^{1/3}$ cm$^2$s$^{-1}$, i.e, 2 - 3 orders of magnitude smaller than the typical diffusion coefficient of the ISM), with a large spatial extent (r$_{halo}$ $\approx$ 20 - 50 pc) \citep{abey17a, ruo22}. CR self-generated turbulence or Alfv\'{e}n waves is popularly considered to be the origin of the slow isotropic diffusion, where a large density gradient of escaped electron-positron pairs near the source induces the growth of small-scale magnetohydrodynamic (MHD) turbulence of the background plasma, otherwise known as the resonant streaming instability. Escaped pairs get trapped by the increased MHD turbulence, which translates into the suppression of the diffusion coefficient. For a comprehensive review, please see \cite{fang22, ruo22} and references therein. Apart from this, multiple models have been proposed to explain the possible origin of TeV halos; namely isotropic, unsuppressed diffusion with the transition from quasi-ballistic propagation \citep{prosekin15}, anisotropic diffusion \citep{ruo19}, etc. Further details regarding the origin of the TeV halo are beyond the scope of this paper. Additionally, the magnetic field associated with the TeV halo was also estimated to be at the same level as the average Galactic magnetic field \citep{sudoh19}, which is quite low compared to that observed in PWNe. From X-ray observations, the magnetic field inside the TeV halo of Geminga was constrained to be $<$ 1 $\mu$G \citep{ruo19b}. Thus, a low estimated magnetic field can also be an important differentiator between the TeV halo and PWN scenarios. 

The presence of a putative GeV pulsar 4FGL J0340.4+5302 co-spatial with the LHAASO source region, and the spatially extended gamma-ray emission observed by LHAASO, hint towards the existence of an extended TeV halo emission in the source region. Although it is difficult to ascertain due to the lack of proper distance estimation, in this work, we assume that the putative pulsar 4FGL J0340.4+5302 is associated with the old, invisible SNR, which makes the age of the pulsar to be $\sim$ 6.2 $\times$ 10$^5$ years. From the non-detection of the old SNR and the offset between the LHAASO source centroid and the 4FGL source, it can be posited that the system is old enough to be in the final stage of evolution, where the host SNR has faded away and the corresponding pulsar has been displaced from its original position due to its natal kick velocity \citep{gaensler06}, which makes the TeV halo scenario more plausible.

Consequently, we have considered a steady-state relativistic electron population from a putative TeV halo associated with the GeV pulsar and calculated the total leptonic contribution from this source to help explain the MWL SED of the LHAASO source. As a result of slow diffusion inside the TeV halo region, radiative cooling timescales of E$_e$ $>$ 10 TeV leptons that produce TeV gamma rays, i.e., $\sim$ 10$^4$(B/10 $\mu$G)$^{-2}$ (E$_e$/10 TeV)$^{-1}$ years \citep{giacinti20}, are comparatively lower than the escape timescale, i.e., $\sim$ 4.4 $\times$ 10$^4$$\left(\frac{r_{halo}}{35\:pc}\right)^2$$\left(\frac{D_0}{4.5 \times 10^{27}\:cm^2 s^{-1}}\right)^{-1}$$\left(\frac{E_{e}}{10\:TeV}\right)^{-\frac{1}{3}}$ years \citep{ruo22}. So, we neglect the effect of lepton escape from the TeV halo source. In a radiation-dominated environment, the inverse-Compton (IC) emission from the accelerated leptons with a hard spectrum, that escape from the disrupted PWN into the TeV halo, can provide a significant contribution to the VHE-UHE gamma-ray regime \citep{breuhaus21}. We have considered different leptonic cooling mechanisms, such as IC and synchrotron \citep{baring, ghisellini, blumenthal}, to obtain the MWL emission from the parent electron population associated with the TeV halo using \texttt{GAMERA} \citep{hahn}. The synchrotron emission, which is constrained by the X-ray upper limit, should also provide a constraint on the value of the associated magnetic field, which would, in turn, confirm the TeV halo interpretation of the observed VHE-UHE gamma-ray emission.

To explain the MWL SED of LHAASO J0341+5258, in this paper, we have considered two scenarios: (a) Two-zone Lepto-hadronic scenario, where TeV halo emission has been used in conjunction with the hadronic emission from SNR+MC association (see discussion in Section \ref{snr}), and (b) One-zone Leptonic scenario, in which the entire emission is explained by an individual TeV halo, without the presence of any SNR+MC association. We have considered the distance of the TeV halo in both cases to be d = 1 kpc. The spectrum of the electron population was assumed to be a simple power law with an exponential cutoff in the forms of N$_{LH}$ $\propto$ E$_e^{-p_{LH}}$ exp(-E$_e$/E$^{max}_{LH}$) for the Lepto-hadronic case, and N$_{L}$ $\propto$ E$_e^{-p_{L}}$ exp(-E$_e$/E$^{max}_{L}$) for the Leptonic case. In this case, E$^{max}_{LH}$ and E$^{max}_{L}$ depict the maximum energy, beyond which the rollover in the spectrum ensues. It can also be portrayed as the rollover energy or the cutoff energy of the spectrum. The minimum energy of the electron population was given by the rest mass energy. Interstellar Radiation Field has been considered following \cite{popescu}, and the associated magnetic field in the two cases has been fixed by remaining consistent with the X-ray upper limits reported in \cite{cao21c}.

In both of the cases, the spectral index of the lepton population was fixed at p$_{LH}$ = p$_{L}$ = 1.5 \citep{sudoh19}. For the Lepto-hadronic case, the maximum energy and the energy budget required to explain the MWL SED are E$^{max}_{LH}$ $\sim$ 60 TeV, and W$_{LH}$ $\sim$ 1.5 $\times$ 10$^{45}$ erg, whereas the same for the Leptonic case were found out to be E$^{max}_{L}$ $\sim$ 120 TeV, and W$_{L}$ $\sim$ 1.7 $\times$ 10$^{45}$ erg. The maximum energy estimates in both of the cases are consistent with the TeV halo scenario, where electrons, with maximum energy ranging from tens to hundreds of TeVs, can be present in the halo region \citep{ruo22}. The associated magnetic fields, which are constrained by the X-ray upper limits, were estimated to be B$_{LH}$ $\approx$ 4 $\mu$G for the Lepto-hadronic case, and B$_{L}$ $\approx$ 2.6 $\mu$G for the Leptonic case. In both of the cases, the values of the estimated magnetic fields are well below that typically observed in a standard PWN and are similar to the average value of the Galactic magnetic field in the ISM (2 - 6 $\mu$G), which corroborates with the TeV halo interpretation of gamma-ray emission. The model spectrum for (a) Lepto-hadronic and (b) Leptonic cases, along with the data points for the MWL SED of LHAASO 0341+5258 taken from \cite{cao21c}, are shown in panels (a) and (b) of Figure \ref{fig2}, respectively. As can be seen from the figures, both cases are consistent with the MWL SED and upper limits hitherto obtained.

\section{Discussion}\label{sec:discussion}

In this section, we discuss the main implications of this work in detail. Since both the Lepto-hadronic and Leptonic models explain the MWL SED of the LHAASO source, it is difficult to distinguish whether the SNR+MC association or the TeV halo is responsible for the UHE gamma-ray emission observed by LHAASO. Due to poor angular resolution capability, LHAASO cannot discern the associated PeVatron in the source region. Consequently, VHE gamma-ray observations are required to properly confirm the source contribution from the study of spatial morphology. From Figure \ref{fig2}, it can be seen that the model spectrum, in both cases, exceeds the sensitivities of VHE gamma-ray observatories such as CTA north \citep{CTA19}, SWGO \citep{SWGO19} and ASTRI \citep{astri23}. Thus, VHE gamma-ray data obtained by these observatories would be crucial to unveil the nature of the PeVatron and confirm which of these two cases is valid. For example, if the entire emission is due to the leptonic component from a TeV halo, then only a singular emission peak should be observed. On the other hand, if the Lepto-hadronic case is valid, then double peaked significance map should be observed in the source region, as it was observed in the case of LHAASO J1908+0621 \citep{desarkar22a,li21}. Hence, from the study of the spatial morphology using VHE gamma-ray data, it will be possible to confirm the nature of the associated PeVatron in this case.
      
Although the point-like nature and a curved SED, explained by the synchro-curvature emission, indicate that 4FGL J0340.4+5302 is likely a GeV pulsar, further observations are needed for its confirmation. A blind search for pulsation or periodicity from this source was not possible without an updated ephemeris. Nevertheless, detection of this putative pulsar in radio wavelength would provide us with information necessary for producing the corresponding ephemeris, which can be used to discover periodicity in the 4FGL source. This conclusion was echoed in the recently published Third Fermi Large Area Telescope Catalog of Gamma-ray Pulsars \citep{smith23}. Although no significant variability was observed with 4FGL J0340.2+5302 (variability index 10.45, which is less than the threshold of 24.7), it is one of four sources with TS $>$ 200, undetected beyond 10 GeV, significantly curved (well fit with a LogParabola function), localization ellipse semi-major axes with 95$\%$ confidence limit $<$ 10$'$, Galactic latitude |b| < 10$^{\circ}$, all of which indicates that this source is suitable for radio searches, and its origin as being a young, energetic pulsar is favorable. The authors mention that radio pulsations from this source will confirm its pulsar origin, but none have been reported to date. Moreover, electron population accelerated in the shock front can also produce HE gamma rays, which might be obscured by the GeV pulsar emission, similar to that observed in LHAASO J1908+0621 \citep{desarkar22a}. Such leptonic emission was also observed in the case of LHAASO J2108+5157 \citep{desarkar23}. Off-pulse analyses of the putative GeV pulsar, using the updated ephemeris, can be performed to uncover previously undetected emissions from the source region in the HE gamma-ray range (see \cite{li21}). 

In Section \ref{sec:synchro}, we have used the synchro-curvature model to explain the SED of 4FGL J0340.4+5302. Due to a lack of knowledge regarding the timing properties of the 4FGL source, e.g. the spin period, some of the model parameters (e.g., r$_c$, B) were fixed at values consistent with \cite{vigano15}. Since the predicted age of the putative GeV pulsar ($\sim$ 6.2 $\times$ 10$^5$ years) is close to that of Geminga's ($\sim$ 3 $\times$ 10$^5$ years), we try to test the consistency of our model by associating the typical parameters of Geminga to the presumptive pulsar discussed in this work. Geminga is a relatively old pulsar with spin period P = 0.237 s, $\Dot{P}$ = 1.0975 $\times$ 10$^{-14}$ \citep{taylor93} and surface magnetic field B$_{Geminga}$ = 3.3 $\times$ 10$^{12}$ G \citep{vigano15b}. For this choice of the spin period, the radius of the light cylinder can be calculated to be r$_{LC}$ = $\frac{P c}{2\pi}$ $\approx$ 1 $\times$ 10$^{19}$ cm. As is usually supposed, the radius of curvature is half of the radius of the light cylinder, i.e., in this case, r$_c$ $\sim$ 5 $\times$ 10$^8$ cm. Accordingly, in the outer magnetosphere of the pulsar, where ultra-relativistic electrons/positrons emit GeV photons via the synchro-curvature process, the magnetic field strength will become B $\sim$ 10$^3$ G. We use these typical values of Geminga in the synchro-curvature model discussed in Section \ref{sec:synchro}, and subsequently try to explain the SED of 4FGL J0340.4+5302. We find that the required values of the free parameters, in this case, come out to be log(E$_{||}$ (V m$^{-1}$)) = 6.740, x$_0$/r$_c$ = 0.07, and N$_0$ = 2 $\times$ 10$^{32}$ \textit{particles}. One can see these new values of the free parameters, compatible with the Geminga-like case, are well within the allowed range of parameter values discussed in Section \ref{sec:synchro}. We further compare the particle number density in this case, with the Goldreich-Julian density limit \citep{goldreich69}, which gives the lower limit of the plasma density in the neutron star magnetosphere. The Goldreich-Julian (GJ) particle number density, given by n$_{GJ}$ = 7 $\times$ 10$^{-2}$ (B$_z$/P) $\textit{particles}$ cm$^{-3}$, depends on the pulsar spin period, magnetic field, and the alignment of the pulsar spin axis with respect to the magnetic field lines \citep{goldreich69}. We use B$_z$ = B$_{Geminga}$, assuming that near the pulsar surface, the spin axis is essentially aligned with the magnetic field lines, which indicates that the corresponding GJ particle number density is n$_{GJ}$ $\approx$ 1 $\times$ 10$^{12}$ \textit{particles} cm$^{-3}$, considering the spin period P of Geminga. The particle number density, in practical cases, should be comparable with or can even greatly exceed n$_{GJ}$ (please see \cite{lyutikov06} and references therein), since n$_{GJ}$ essentially indicates the uncompensated charges in the region. With that in mind, we calculate the particle number density for the Geminga-like case discussed above and compare it with the GJ density. The total effective number of particles has been calculated by integrating equation \ref{eq16a}, assuming a spherical emission volume with a radius of 10$^6$ cm. For the total number of charged particles in that emission volume, N$_e$ $\approx$ 5.6 $\times$ 10$^{30}$ \textit{particles}, we find that the corresponding particle number density comes out to be n$_e$ $\approx$ 1.3 $\times$ 10$^{12}$ \textit{particles} cm$^{-3}$. So, assuming an emission volume similar to that of the pulsar, the particle number density of the model (n$_e$) is found to be comparable with the theoretical expectation provided by the GJ limit (n$_{GJ}$), which reflects the consistency of the model. Note that the particle number is dependent on the position (as indicated by equation \ref{eq16a}), and if a larger emission volume is considered, n$_e$ will be much smaller than that estimated above. However, the magnetic field will also decrease drastically, away from the surface of the pulsar, which means the condition n$_e$ $\geq$ n$_{GJ}$ will continue to hold, even if it is considered far away from the surface. Since there are uncertainties regarding the distance, spin period, and magnetic field of the putative pulsar, we only aim to provide rough estimates to show the consistency of the synchro-curvature model, when Geminga-like parameters are assumed. Future observations, especially in the radio wavelengths, confirming these unknown variables, will help solidify the pulsar origin of the 4FGL source.
      
Finally, radio observations of the source region are necessary to constrain the synchrotron emission from the TeV halo. Accelerated electron population that got injected inside the MCs can also produce synchrotron emission when interacting with the very high magnetic field inside the MCs \citep{desarkar22a,desarkar23}. Radio upper limits from further observations can also constrain the leptonic contribution from the SNR+MC association.
   
\section{Conclusion}\label{sec:conclusion}

In this paper, we have discussed the nature and emission of UHE gamma-ray source LHAASO J0341+5258 in a MWL context. Future studies, taking into account the appropriate distance corresponding to each source, may provide better constraints to the considered model parameters. Nevertheless, the MWL SED observed to date can be satisfactorily explained by both Lepto-hadronic and Leptonic models considered in this work. Moreover, we have consistently shown that the GeV counterpart of the LHAASO source, 4FGL J0340.4+5302, is likely a GeV pulsar. Furthermore, we have also discussed the implications of our model and provided justifications for further observations in multiple wavelengths, which are necessary to confirm the source association and radiation mechanism associated with this enigmatic source.

\begin{acknowledgements}
The authors thank the anonymous reviewer for useful suggestions and constructive criticism. ADS thanks Shiv Sethi for the useful discussions.     
\end{acknowledgements}

%
\bibliographystyle{aa} 
\bibliography{aa.bib} 

\begin{thebibliography}{56}
\expandafter\ifx\csname natexlab\endcsname\relax\def\natexlab#1{#1}\fi

\bibitem[{{Aartsen} {et~al.}(2021){Aartsen}, {Abbasi}, {Ackermann}, {Adams},
  {Aguilar}, {Ahlers}, {Ahrens}, {Alispach}, {Allison}, {Amin}, {Andeen},
  {Anderson}, {Ansseau}, {Anton}, {Arg{\"u}elles}, {Arlen}, {Auffenberg},
  {Axani}, {Bagherpour}, {Bai}, {Balagopal V}, {Barbano}, {Bartos}, {Bastian},
  {Basu}, {Baum}, {Baur}, {Bay}, {Beatty}, {Becker}, {Tjus}, {BenZvi},
  {Berley}, {Bernardini}, {Besson}, {Binder}, {Bindig}, {Blaufuss}, {Blot},
  {Bohm}, {Bohmer}, {B{\"o}ser}, {Botner}, {B{\"o}ttcher}, {Bourbeau},
  {Bourbeau}, {Bradascio}, {Braun}, {Bron}, {Brostean-Kaiser}, {Burgman},
  {Burley}, {Buscher}, {Busse}, {Bustamante}, {Campana}, {Carnie-Bronca},
  {Carver}, {Chen}, {Chen}, {Cheung}, {Chirkin}, {Choi}, {Clark}, {Clark},
  {Classen}, {Coleman}, {Collin}, {Connolly}, {Conrad}, {Coppin}, {Correa},
  {Cowen}, {Cross}, {Dave}, {Deaconu}, {De Clercq}, {DeLaunay}, {De Kockere},
  {Dembinski}, {Deoskar}, {De Ridder}, {Desai}, {Desiati}, {de Vries}, {de
  Wasseige}, {de With}, {DeYoung}, {Dharani}, {Diaz}, {D{\'\i}az-V{\'e}lez},
  {Dujmovic}, {Dunkman}, {DuVernois}, {Dvorak}, {Ehrhardt}, {Eller}, {Engel},
  {Evans}, {Evenson}, {Fahey}, {Farrag}, {Fazely}, {Felde}, {Fienberg},
  {Filimonov}, {Finley}, {Fischer}, {Fox}, {Franckowiak}, {Friedman}, {Fritz},
  {Gaisser}, {Gallagher}, {Ganster}, {Garcia-Fernandez}, {Garrappa}, {Gartner},
  {Gerhard}, {Gernhaeuser}, {Ghadimi}, {Glaser}, {Glauch}, {Gl{\"u}senkamp},
  {Goldschmidt}, {Gonzalez}, {Goswami}, {Grant}, {Gr{\'e}goire}, {Griffith},
  {Griswold}, {G{\"u}nd{\"u}z}, {Haack}, {Hallgren}, {Halliday}, {Halve},
  {Halzen}, {Hanson}, {Hanson}, {Hardin}, {Haugen}, {Haungs}, {Hauser},
  {Hebecker}, {Heinen}, {Heix}, {Helbing}, {Hellauer}, {Henningsen},
  {Hickford}, {Hignight}, {Hill}, {Hill}, {Hoffman}, {Hoffmann}, {Hoffmann},
  {Hoinka}, {Hokanson-Fasig}, {Holzapfel}, {Hoshina}, {Huang}, {Huber},
  {Huber}, {Huege}, {Hughes}, {Hultqvist}, {H{\"u}nnefeld}, {Hussain}, {In},
  {Iovine}, {Ishihara}, {Jansson}, {Japaridze}, {Jeong}, {Jones}, {Jonske},
  {Joppe}, {Kalekin}, {Kang}, {Kang}, {Kang}, {Kappes}, {Kappesser}, {Karg},
  {Karl}, {Karle}, {Katori}, {Katz}, {Kauer}, {Keivani}, {Kellermann},
  {Kelley}, {Kheirandish}, {Kim}, {Kin}, {Kintscher}, {Kiryluk}, {Kittler},
  {Kleifges}, {Klein}, {Koirala}, {Kolanoski}, {K{\"o}pke}, {Kopper}, {Kopper},
  {Koskinen}, {Koundal}, {Kovacevich}, {Kowalski}, {Krauss}, {Krings},
  {Kr{\"u}ckl}, {Kulacz}, {Kurahashi}, {Gualda}, {Lahmann}, {Lanfranchi},
  {Larson}, {Latif}, {Lauber}, {Lazar}, {Leonard}, {Leszczy{\'n}ska}, {Li},
  {Liu}, {Lohfink}, {LoSecco}, {Mariscal}, {Lu}, {Lucarelli}, {Ludwig},
  {L{\"u}nemann}, {Luszczak}, {Lyu}, {Ma}, {Madsen}, {Maggi}, {Mahn}, {Makino},
  {Mallik}, {Mancina}, {Mandalia}, {Mari{\c{s}}}, {Marka}, {Marka}, {Maruyama},
  {Mase}, {Maunu}, {McNally}, {Meagher}, {Medina}, {Meier}, {Meighen-Berger},
  {Merz}, {Meyers}, {Micallef}, {Mockler}, {Moment{\'e}}, {Montaruli}, {Moore},
  {Morse}, {Moulai}, {Muth}, {Naab}, {Nagai}, {Nam}, {Nauman}, {Necker},
  {Neer}, {Nelles}, {Nguyễn}, {Niederhausen}, {Nisa}, {Nowicki}, {Nygren},
  {Oberla}, {Pollmann}, {Oehler}, {Olivas}, {O'Sullivan}, {Pan}, {Pandya},
  {Pankova}, {Papp}, {Park}, {Parker}, {Paudel}, {Peiffer}, {P{\'e}rez de los
  Heros}, {Petersen}, {Philippen}, {Pieloth}, {Pieper}, {Pinfold}, {Pizzuto},
  {Plaisier}, {Plum}, {Popovych}, {Porcelli}, {Rodriguez}, {Price},
  {Przybylski}, {Raab}, {Raissi}, {Rameez}, {Rauch}, {Rawlins}, {Rea},
  {Rehman}, {Reimann}, {Renschler}, {Renzi}, {Resconi}, {Reusch}, {Rhode},
  {Richman}, {Riedel}, {Riegel}, {Roberts}, {Robertson}, {Roellinghoff},
  {Rongen}, {Rott}, {Ruhe}, {Ryckbosch}, {Cantu}, {Safa}, {Herrera},
  {Sandrock}, {Sandroos}, {Sandstrom}, {Santander}, {Sarkar}, {Sarkar},
  {Satalecka}, {Scharf}, {Schaufel}, {Schieler}, {Schlunder}, {Schmidt},
  {Schneider}, {Schneider}, {Schr{\"o}der}, {Schumacher}, {Sclafani}, {Seckel},
  {Seunarine}, {Shaevitz}, {Sharma}, {Shefali}, {Silva}, {Smith}, {Smithers},
  {Snihur}, {Soedingrekso}, {Soldin}, {S{\"o}ldner-Rembold}, {Song},
  {Southall}, {Spiczak}, {Spiering}, {Stachurska}, {Stamatikos}, {Stanev},
  {Stein}, {Stettner}, {Steuer}, {Stezelberger}, {Stokstad}, {Strotjohann},
  {St{\"u}rwald}, {Stuttard}, {Sullivan}, {Taboada}, {Taketa}, {Tanaka},
  {Tenholt}, {Ter-Antonyan}, {Terliuk}, {Tilav}, {Tollefson}, {Tomankova},
  {T{\"o}nnis}, {Torres}, {Toscano}, {Tosi}, {Trettin}, {Tselengidou}, {Tung},
  {Turcati}, {Turcotte}, {Turley}, {Twagirayezu}, {Ty}, {Unger}, {Elorrieta},
  {Vandenbroucke}, {van Eijk}, {van Eijndhoven}, {Vannerom}, {van Santen},
  {Veberic}, {Verpoest}, {Vieregg}, {Vraeghe}, {Walck}, {Watson}, {Weaver},
  {Weindl}, {Weinstock}, {Weiss}, {Weldert}, {Welling}, {Wendt}, {Werthebach},
  {Whitehorn}, {Wiebe}, {Wiebusch}, {Williams}, {Wissel}, {Wolf}, {Wood},
  {Woschnagg}, {Wrede}, {Wren}, {Wulff}, {Xu}, {Xu}, {Yanez}, {Yoshida},
  {Yuan}, {Zhang}, {Zierke}, \& {Z{\"o}cklein}}]{aartsen21}
{Aartsen}, M.~G., {Abbasi}, R., {Ackermann}, M., {et~al.} 2021, Journal of
  Physics G Nuclear Physics, 48, 060501

\bibitem[{{Abdo} {et~al.}(2009){Abdo}, {Allen}, {Aune}, {Berley}, {Chen},
  {Christopher}, {DeYoung}, {Dingus}, {Ellsworth}, {Gonzalez}, {Goodman},
  {Hays}, {Hoffman}, {H{\"u}ntemeyer}, {Kolterman}, {Linnemann}, {McEnery},
  {Morgan}, {Mincer}, {Nemethy}, {Pretz}, {Ryan}, {Saz Parkinson}, {Shoup},
  {Sinnis}, {Smith}, {Vasileiou}, {Walker}, {Williams}, \& {Yodh}}]{milagro09}
{Abdo}, A.~A., {Allen}, B.~T., {Aune}, T., {et~al.} 2009, \apjl, 700, L127

\bibitem[{{Abdollahi} {et~al.}(2022){Abdollahi}, {Acero}, {Baldini}, {Ballet},
  {Bastieri}, {Bellazzini}, {Berenji}, {Berretta}, {Bissaldi}, {Blandford},
  {Bloom}, {Bonino}, {Brill}, {Britto}, {Bruel}, {Burnett}, {Buson}, {Cameron},
  {Caputo}, {Caraveo}, {Castro}, {Chaty}, {Cheung}, {Chiaro}, {Cibrario},
  {Ciprini}, {Coronado-Bl{\'a}zquez}, {Crnogorcevic}, {Cutini}, {D'Ammando},
  {De Gaetano}, {Digel}, {Di Lalla}, {Dirirsa}, {Di Venere}, {Dom{\'\i}nguez},
  {Fallah Ramazani}, {Fegan}, {Ferrara}, {Fiori}, {Fleischhack}, {Franckowiak},
  {Fukazawa}, {Funk}, {Fusco}, {Galanti}, {Gammaldi}, {Gargano}, {Garrappa},
  {Gasparrini}, {Giacchino}, {Giglietto}, {Giordano}, {Giroletti}, {Glanzman},
  {Green}, {Grenier}, {Grondin}, {Guillemot}, {Guiriec}, {Gustafsson},
  {Harding}, {Hays}, {Hewitt}, {Horan}, {Hou}, {J{\'o}hannesson}, {Karwin},
  {Kayanoki}, {Kerr}, {Kuss}, {Landriu}, {Larsson}, {Latronico},
  {Lemoine-Goumard}, {Li}, {Liodakis}, {Longo}, {Loparco}, {Lott}, {Lubrano},
  {Maldera}, {Malyshev}, {Manfreda}, {Mart{\'\i}-Devesa}, {Mazziotta}, {Mereu},
  {Meyer}, {Michelson}, {Mirabal}, {Mitthumsiri}, {Mizuno}, {Moiseev},
  {Monzani}, {Morselli}, {Moskalenko}, {Negro}, {Nuss}, {Omodei}, {Orienti},
  {Orlando}, {Paneque}, {Pei}, {Perkins}, {Persic}, {Pesce-Rollins},
  {Petrosian}, {Pillera}, {Poon}, {Porter}, {Principe}, {Rain{\`o}}, {Rando},
  {Rani}, {Razzano}, {Razzaque}, {Reimer}, {Reimer}, {Reposeur},
  {S{\'a}nchez-Conde}, {Saz Parkinson}, {Scotton}, {Serini}, {Sgr{\`o}},
  {Siskind}, {Smith}, {Spandre}, {Spinelli}, {Sueoka}, {Suson}, {Tajima},
  {Tak}, {Thayer}, {Thompson}, {Torres}, {Troja}, {Valverde}, {Wood}, \&
  {Zaharijas}}]{abdol22}
{Abdollahi}, S., {Acero}, F., {Baldini}, L., {et~al.} 2022, \apjs, 260, 53

\bibitem[{{Abe} {et~al.}(2023){Abe}, {Abe}, {Acciari}, {Agudo}, {Aniello},
  {Ansoldi}, {Antonelli}, {Arbet Engels}, {Arcaro}, {Artero}, {Asano}, {Baack},
  {Babi{\'c}}, {Baquero}, {Barres de Almeida}, {Barrio}, {Batkovi{\'c}},
  {Baxter}, {Becerra Gonz{\'a}lez}, {Bednarek}, {Bernardini}, {Bernardos},
  {Berti}, {Besenrieder}, {Bhattacharyya}, {Bigongiari}, {Biland}, {Blanch},
  {Bonnoli}, {Bo{\v{s}}njak}, {Burelli}, {Busetto}, {Carosi},
  {Carretero-Castrillo}, {Castro-Tirado}, {Ceribella}, {Chai}, {Chilingarian},
  {Cikota}, {Colombo}, {Contreras}, {Cortina}, {Covino}, {D'Amico}, {D'Elia},
  {da Vela}, {Dazzi}, {de Angelis}, {de Lotto}, {Del Popolo}, {Delfino},
  {Delgado}, {Delgado Mendez}, {Depaoli}, {di Pierro}, {di Venere}, {Do Souto
  Espi{\~n}eira}, {Dominis Prester}, {Donini}, {Dorner}, {Doro}, {Elsaesser},
  {Emery}, {Escudero}, {Fallah Ramazani}, {Fari{\~n}a}, {Fattorini}, {Font},
  {Fruck}, {Fukami}, {Fukazawa}, {Garc{\'\i}a L{\'o}pez}, {Garczarczyk},
  {Gasparyan}, {Gaug}, {Giesbrecht Paiva}, {Giglietto}, {Giordano}, {Gliwny},
  {Godinovi{\'c}}, {Grau}, {Green}, {Green}, {Hadasch}, {Hahn}, {Hassan},
  {Heckmann}, {Herrera}, {Hrupec}, {H{\"u}tten}, {Imazawa}, {Inada}, {Iotov},
  {Ishio}, {Jim{\'e}nez Mart{\'\i}nez}, {Jormanainen}, {Kerszberg},
  {Kobayashi}, {Kubo}, {Kushida}, {Lamastra}, {Lelas}, {Leone}, {Lindfors},
  {Linhoff}, {Lombardi}, {Longo}, {L{\'o}pez-Coto}, {L{\'o}pez-Moya},
  {L{\'o}pez-Oramas}, {Loporchio}, {Lorini}, {Lyard}, {Machado de Oliveira
  Fraga}, {Majumdar}, {Makariev}, {Maneva}, {Mang}, {Manganaro}, {Mangano},
  {Mannheim}, {Mariotti}, {Mart{\'\i}nez}, {Mas Aguilar}, {Mazin}, {Menchiari},
  {Mender}, {Mi{\'c}anovi{\'c}}, {Miceli}, {Miener}, {Miranda}, {Mirzoyan},
  {Molina}, {Mondal}, {Moralejo}, {Morcuende}, {Moreno}, {Nakamori}, {Nanci},
  {Nava}, {Neustroev}, {Nievas Rosillo}, {Nigro}, {Nilsson}, {Nishijima}, {Njoh
  Ekoume}, {Noda}, {Nozaki}, {Ohtani}, {Oka}, {Okumura}, {Otero-Santos},
  {Paiano}, {Palatiello}, {Paneque}, {Paoletti}, {Paredes}, {Pavleti{\'c}},
  {Persic}, {Pihet}, {Pirola}, {Podobnik}, {Prada Moroni}, {Prandini},
  {Principe}, {Priyadarshi}, {Rhode}, {Rib{\'o}}, {Rico}, {Righi},
  {Rugliancich}, {Sahakyan}, {Saito}, {Sakurai}, {Satalecka}, {Saturni},
  {Schleicher}, {Schmidt}, {Schmuckermaier}, {Schubert}, {Schweizer},
  {Sitarek}, {Sliusar}, {Sobczynska}, {Spolon}, {Stamerra},
  {Stri{\v{s}}kovi{\'c}}, {Strom}, {Strzys}, {Suda}, {Suri{\'c}}, {Tajima},
  {Takahashi}, {Takeishi}, {Tavecchio}, {Temnikov}, {Terauchi}, {Terzi{\'c}},
  {Teshima}, {Tosti}, {Truzzi}, {Tutone}, {Ubach}, {van Scherpenberg}, {Vazquez
  Acosta}, {Ventura}, {Verguilov}, {Viale}, {Vigorito}, {Vitale}, {Vovk},
  {Walter}, {Will}, {Wunderlich}, {Yamamoto}, \& {Zari{\'c}}}]{abe23}
{Abe}, H., {Abe}, S., {Acciari}, V.~A., {et~al.} 2023, \aap, 671, A12

\bibitem[{{Abeysekara} {et~al.}(2017{\natexlab{a}}){Abeysekara}, {Albert},
  {Alfaro}, {Alvarez}, {{\'A}lvarez}, {Arceo}, {Arteaga-Vel{\'a}zquez}, {Avila
  Rojas}, {Ayala Solares}, {Barber}, {Bautista-Elivar}, {Becerril},
  {Belmont-Moreno}, {BenZvi}, {Berley}, {Bernal}, {Braun}, {Brisbois},
  {Caballero-Mora}, {Capistr{\'a}n}, {Carrami{\~n}ana}, {Casanova}, {Castillo},
  {Cotti}, {Cotzomi}, {Couti{\~n}o de Le{\'o}n}, {De Le{\'o}n}, {De la Fuente},
  {Dingus}, {DuVernois}, {D{\'\i}az-V{\'e}lez}, {Ellsworth}, {Engel},
  {Enr{\'\i}quez-Rivera}, {Fiorino}, {Fraija}, {Garc{\'\i}a-Gonz{\'a}lez},
  {Garfias}, {Gerhardt}, {Gonz{\'a}lez Mu{\~n}oz}, {Gonz{\'a}lez}, {Goodman},
  {Hampel-Arias}, {Harding}, {Hern{\'a}ndez}, {Hern{\'a}ndez-Almada}, {Hinton},
  {Hona}, {Hui}, {H{\"u}ntemeyer}, {Iriarte}, {Jardin-Blicq}, {Joshi},
  {Kaufmann}, {Kieda}, {Lara}, {Lauer}, {Lee}, {Lennarz}, {Vargas},
  {Linnemann}, {Longinotti}, {Luis Raya}, {Luna-Garc{\'\i}a}, {L{\'o}pez-Coto},
  {Malone}, {Marinelli}, {Martinez}, {Martinez-Castellanos},
  {Mart{\'\i}nez-Castro}, {Mart{\'\i}nez-Huerta}, {Matthews},
  {Miranda-Romagnoli}, {Moreno}, {Mostaf{\'a}}, {Nellen}, {Newbold}, {Nisa},
  {Noriega-Papaqui}, {Pelayo}, {Pretz}, {P{\'e}rez-P{\'e}rez}, {Ren}, {Rho},
  {Rivi{\`e}re}, {Rosa-Gonz{\'a}lez}, {Rosenberg}, {Ruiz-Velasco}, {Salazar},
  {Salesa Greus}, {Sandoval}, {Schneider}, {Schoorlemmer}, {Sinnis}, {Smith},
  {Springer}, {Surajbali}, {Taboada}, {Tibolla}, {Tollefson}, {Torres},
  {Ukwatta}, {Vianello}, {Weisgarber}, {Westerhoff}, {Wisher}, {Wood},
  {Yapici}, {Yodh}, {Younk}, {Zepeda}, {Zhou}, {Guo}, {Hahn}, {Li}, \&
  {Zhang}}]{abey17a}
{Abeysekara}, A.~U., {Albert}, A., {Alfaro}, R., {et~al.} 2017{\natexlab{a}},
  Science, 358, 911

\bibitem[{{Abeysekara} {et~al.}(2017{\natexlab{b}}){Abeysekara}, {Albert},
  {Alfaro}, {Alvarez}, {{\'A}lvarez}, {Arceo}, {Arteaga-Vel{\'a}zquez}, {Ayala
  Solares}, {Barber}, {Baughman}, {Bautista-Elivar}, {Becerra Gonzalez},
  {Becerril}, {Belmont-Moreno}, {BenZvi}, {Berley}, {Bernal}, {Braun},
  {Brisbois}, {Caballero-Mora}, {Capistr{\'a}n}, {Carrami{\~n}ana}, {Casanova},
  {Castillo}, {Cotti}, {Cotzomi}, {Couti{\~n}o de Le{\'o}n}, {de la Fuente},
  {De Le{\'o}n}, {Diaz Hernandez}, {Dingus}, {DuVernois},
  {D{\'\i}az-V{\'e}lez}, {Ellsworth}, {Engel}, {Fiorino}, {Fraija},
  {Garc{\'\i}a-Gonz{\'a}lez}, {Garfias}, {Gerhardt}, {Gonz{\'a}lez Mu{\~n}oz},
  {Gonz{\'a}lez}, {Goodman}, {Hampel-Arias}, {Harding}, {Hernandez},
  {Hernandez-Almada}, {Hinton}, {Hui}, {H{\"u}ntemeyer}, {Iriarte},
  {Jardin-Blicq}, {Joshi}, {Kaufmann}, {Kieda}, {Lara}, {Lauer}, {Lee},
  {Lennarz}, {Le{\'o}n Vargas}, {Linnemann}, {Longinotti}, {Raya},
  {Luna-Garc{\'\i}a}, {L{\'o}pez-Coto}, {Malone}, {Marinelli}, {Martinez},
  {Martinez-Castellanos}, {Mart{\'\i}nez-Castro}, {Mart{\'\i}nez-Huerta},
  {Matthews}, {Miranda-Romagnoli}, {Moreno}, {Mostaf{\'a}}, {Nellen},
  {Newbold}, {Nisa}, {Noriega-Papaqui}, {Pelayo}, {Pretz},
  {P{\'e}rez-P{\'e}rez}, {Ren}, {Rho}, {Rivi{\`e}re}, {Rosa-Gonz{\'a}lez},
  {Rosenberg}, {Ruiz-Velasco}, {Salazar}, {Salesa Greus}, {Sandoval},
  {Schneider}, {Schoorlemmer}, {Sinnis}, {Smith}, {Springer}, {Surajbali},
  {Taboada}, {Tibolla}, {Tollefson}, {Torres}, {Ukwatta}, {Vianello},
  {Villase{\~n}or}, {Weisgarber}, {Westerhoff}, {Wisher}, {Wood}, {Yapici},
  {Younk}, {Zepeda}, \& {Zhou}}]{abey17b}
{Abeysekara}, A.~U., {Albert}, A., {Alfaro}, R., {et~al.} 2017{\natexlab{b}},
  \apj, 843, 40

\bibitem[{{Abeysekara} {et~al.}(2020){Abeysekara}, {Albert}, {Alfaro}, {Angeles
  Camacho}, {Arteaga-Vel{\'a}zquez}, {Arunbabu}, {Avila Rojas}, {Ayala
  Solares}, {Baghmanyan}, {Belmont-Moreno}, {BenZvi}, {Brisbois},
  {Caballero-Mora}, {Capistr{\'a}n}, {Carrami{\~n}ana}, {Casanova}, {Cotti},
  {Cotzomi}, {Couti{\~n}o de Le{\'o}n}, {De la Fuente}, {de Le{\'o}n},
  {Dichiara}, {Dingus}, {DuVernois}, {D{\'\i}az-V{\'e}lez}, {Ellsworth},
  {Engel}, {Espinoza}, {Fleischhack}, {Fraija}, {Galv{\'a}n-G{\'a}mez},
  {Garcia}, {Garc{\'\i}a-Gonz{\'a}lez}, {Garfias}, {Gonz{\'a}lez}, {Goodman},
  {Harding}, {Hernandez}, {Hinton}, {Hona}, {Huang}, {Hueyotl-Zahuantitla},
  {H{\"u}ntemeyer}, {Iriarte}, {Jardin-Blicq}, {Joshi}, {Kaufmann}, {Kieda},
  {Lara}, {Lee}, {Le{\'o}n Vargas}, {Linnemann}, {Longinotti}, {Luis-Raya},
  {Lundeen}, {L{\'o}pez-Coto}, {Malone}, {Marinelli}, {Martinez},
  {Martinez-Castellanos}, {Mart{\'\i}nez-Castro}, {Mart{\'\i}nez-Huerta},
  {Matthews}, {Miranda-Romagnoli}, {Morales-Soto}, {Moreno}, {Mostaf{\'a}},
  {Nayerhoda}, {Nellen}, {Newbold}, {Nisa}, {Noriega-Papaqui}, {Peisker},
  {P{\'e}rez-P{\'e}rez}, {Pretz}, {Ren}, {Rho}, {Rivi{\`e}re},
  {Rosa-Gonz{\'a}lez}, {Rosenberg}, {Ruiz-Velasco}, {Salesa Greus}, {Sandoval},
  {Schneider}, {Schoorlemmer}, {Sinnis}, {Smith}, {Springer}, {Surajbali},
  {Tabachnick}, {Tanner}, {Tibolla}, {Tollefson}, {Torres}, {Torres-Escobedo},
  {Villase{\~n}or}, {Weisgarber}, {Wood}, {Yapici}, {Zhang}, {Zhou}, \& {HAWC
  Collaboration}}]{abeysekara20}
{Abeysekara}, A.~U., {Albert}, A., {Alfaro}, R., {et~al.} 2020, \prl, 124,
  021102

\bibitem[{{Albert} {et~al.}(2019){Albert}, {Alfaro}, {Ashkar}, {Alvarez},
  {{\'A}lvarez}, {Arteaga-Vel{\'a}zquez}, {Ayala Solares}, {Arceo}, {Bellido},
  {BenZvi}, {Bretz}, {Brisbois}, {Brown}, {Brun}, {Caballero-Mora}, {Carosi},
  {Carrami{\~n}ana}, {Casanova}, {Chadwick}, {Cotter}, {Couti{\~n}o De
  Le{\'o}n}, {Cristofari}, {Dasso}, {de la Fuente}, {Dingus}, {Desiati},
  {Salles}, {de Souza}, {Dorner}, {D{\'\i}az-V{\'e}lez},
  {Garc{\'\i}a-Gonz{\'a}lez}, {DuVernois}, {Di Sciascio}, {Engel},
  {Fleischhack}, {Fraija}, {Funk}, {Glicenstein}, {Gonzalez}, {Gonz{\'a}lez},
  {Goodman}, {Harding}, {Haungs}, {Hinton}, {Hona}, {Hoyos}, {Huentemeyer},
  {Iriarte}, {Jardin-Blicq}, {Joshi}, {Kaufmann}, {Kawata}, {Kunwar},
  {Lefaucheur}, {Lenain}, {Link}, {L{\'o}pez-Coto}, {Marandon}, {Mariotti},
  {Mart{\'\i}nez-Castro}, {Mart{\'\i}nez-Huerta}, {Mostaf{\'a}}, {Nayerhoda},
  {Nellen}, {de O{\~n}a Wilhelmi}, {Parsons}, {Patricelli}, {Pichel}, {Piel},
  {Prandini}, {Pueschel}, {Procureur}, {Reisenegger}, {Rivi{\`e}re},
  {Rodriguez}, {Rovero}, {Rowell}, {Ruiz-Velasco}, {Sandoval}, {Santander},
  {Sako}, {Sako}, {Satalecka}, {Schoorlemmer}, {Sch{\"u}ssler},
  {Seglar-Arroyo}, {Smith}, {Spencer}, {Surajbali}, {Tabachnick}, {Taylor},
  {Tibolla}, {Torres}, {Vallage}, {Viana}, {Watson}, {Weisgarber}, {Werner},
  {White}, {Wischnewski}, {Yang}, {Zepeda}, \& {Zhou}}]{SWGO19}
{Albert}, A., {Alfaro}, R., {Ashkar}, H., {et~al.} 2019, arXiv e-prints,
  arXiv:1902.08429

\bibitem[{{Amenomori} {et~al.}(2019){Amenomori}, {Bao}, {Bi}, {Chen}, {Chen},
  {Chen}, {Chen}, {Chen}, {Cirennima}, {Cui}, {Danzengluobu}, {Ding}, {Fang},
  {Fang}, {Feng}, {Feng}, {Feng}, {Gao}, {Gou}, {Guo}, {He}, {He}, {Hibino},
  {Hotta}, {Hu}, {Hu}, {Huang}, {Jia}, {Jiang}, {Jin}, {Kajino}, {Kasahara},
  {Katayose}, {Kato}, {Kato}, {Kawata}, {Kozai}, {Labaciren}, {Le}, {Li}, {Li},
  {Li}, {Lin}, {Liu}, {Liu}, {Liu}, {Liu}, {Lou}, {Lu}, {Meng}, {Mitsui},
  {Munakata}, {Nakamura}, {Nanjo}, {Nishizawa}, {Ohnishi}, {Ohta}, {Ozawa},
  {Qian}, {Qu}, {Saito}, {Sakata}, {Sako}, {Sengoku}, {Shao}, {Shibata},
  {Shiomi}, {Sugimoto}, {Takita}, {Tan}, {Tateyama}, {Torii}, {Tsuchiya},
  {Udo}, {Wang}, {Wu}, {Xue}, {Yagisawa}, {Yamamoto}, {Yang}, {Yuan}, {Zhai},
  {Zhang}, {Zhang}, {Zhang}, {Zhang}, {Zhang}, {Zhang}, {Zhang},
  {Zhaxisangzhu}, {Zhou}, \& {Tibet AS {\ensuremath{\gamma}}
  Collaboration}}]{amenomori19}
{Amenomori}, M., {Bao}, Y.~W., {Bi}, X.~J., {et~al.} 2019, \prl, 123, 051101

\bibitem[{{Baring} {et~al.}(1999){Baring}, {Ellison}, {Reynolds}, {Grenier}, \&
  {Goret}}]{baring}
{Baring}, M.~G., {Ellison}, D.~C., {Reynolds}, S.~P., {Grenier}, I.~A., \&
  {Goret}, P. 1999, \apj, 513, 311

\bibitem[{{Blumenthal} \& {Gould}(1970)}]{blumenthal}
{Blumenthal}, G.~R. \& {Gould}, R.~J. 1970, Reviews of Modern Physics, 42, 237

\bibitem[{{Breuhaus} {et~al.}(2021){Breuhaus}, {Hahn}, {Romoli}, {Reville},
  {Giacinti}, {Tuffs}, \& {Hinton}}]{breuhaus21}
{Breuhaus}, M., {Hahn}, J., {Romoli}, C., {et~al.} 2021, \apjl, 908, L49

\bibitem[{Cao(2010)}]{lhaaso10}
Cao, Z. 2010, Chinese Physics C, 34, 249

\bibitem[{{Cao} {et~al.}(2021{\natexlab{a}}){Cao}, {Aharonian}, {An},
  {Axikegu}, {Bai}, {Bai}, {Bao}, {Bastieri}, {Bi}, {Bi}, {Cai}, {Cai}, {Cao},
  {Chang}, {Chang}, {Chen}, {Chen}, {Chen}, {Chen}, {Chen}, {Chen}, {Chen},
  {Chen}, {Chen}, {Chen}, {Chen}, {Chen}, {Chen}, {Chen}, {Cheng}, {Cheng},
  {Cui}, {Cui}, {Cui}, {D'Ettorre Piazzoli}, {Dai}, {Dai}, {Dai},
  {Danzengluobu}, {Della Volpe}, {Dong}, {Duan}, {Fan}, {Fan}, {Fan}, {Fang},
  {Fang}, {Feng}, {Feng}, {Feng}, {Feng}, {Gao}, {Gao}, {Gao}, {Gao}, {Gao},
  {Ge}, {Geng}, {Gong}, {Gou}, {Gu}, {Guo}, {Guo}, {Guo}, {Guo}, {Guo}, {Han},
  {He}, {He}, {He}, {He}, {He}, {He}, {Heller}, {Hor}, {Hou}, {Hou}, {Hu},
  {Hu}, {Hu}, {Hu}, {Huang}, {Huang}, {Huang}, {Huang}, {Huang}, {Huang}, {Ji},
  {Ji}, {Jia}, {Jiang}, {Jiang}, {Jin}, {Ke}, {Kuleshov}, {Levochkin}, {Li},
  {Li}, {Li}, {Li}, {Li}, {Li}, {Li}, {Li}, {Li}, {Li}, {Li}, {Li}, {Li}, {Li},
  {Li}, {Li}, {Li}, {Li}, {Liang}, {Liang}, {Lin}, {Liu}, {Liu}, {Liu}, {Liu},
  {Liu}, {Liu}, {Liu}, {Liu}, {Liu}, {Liu}, {Liu}, {Liu}, {Liu}, {Liu}, {Liu},
  {Liu}, {Long}, {Lu}, {Lv}, {Ma}, {Ma}, {Ma}, {Mao}, {Masood}, {Min},
  {Mitthumsiri}, {Montaruli}, {Nan}, {Pang}, {Pattarakijwanich}, {Pei}, {Qi},
  {Qi}, {Qiao}, {Qin}, {Ruffolo}, {Rulev}, {Saiz}, {Shao}, {Shchegolev},
  {Sheng}, {Shi}, {Song}, {Stenkin}, {Stepanov}, {Su}, {Sun}, {Sun}, {Sun},
  {Tam}, {Tang}, {Tian}, {Wang}, {Wang}, {Wang}, {Wang}, {Wang}, {Wang},
  {Wang}, {Wang}, {Wang}, {Wang}, {Wang}, {Wang}, {Wang}, {Wang}, {Wang},
  {Wang}, {Wang}, {Wang}, {Wang}, {Wang}, {Wang}, {Wang}, {Wei}, {Wei}, {Wei},
  {Wen}, {Wu}, {Wu}, {Wu}, {Wu}, {Wu}, {Xi}, {Xia}, {Xia}, {Xiang}, {Xiao},
  {Xiao}, {Xiao}, {Xin}, {Xin}, {Xing}, {Xu}, {Xu}, {Xue}, {Yan}, {Yan},
  {Yang}, {Yang}, {Yang}, {Yang}, {Yang}, {Yang}, {Yang}, {Yao}, {Yao}, {Ye},
  {Yin}, {Yin}, {You}, {You}, {Yu}, {Yuan}, {Zeng}, {Zeng}, {Zeng}, {Zeng},
  {Zha}, {Zhai}, {Zhang}, {Zhang}, {Zhang}, {Zhang}, {Zhang}, {Zhang}, {Zhang},
  {Zhang}, {Zhang}, {Zhang}, {Zhang}, {Zhang}, {Zhang}, {Zhang}, {Zhang},
  {Zhang}, {Zhang}, {Zhang}, {Zhang}, {Zhao}, {Zhao}, {Zhao}, {Zhao}, {Zhao},
  {Zheng}, {Zheng}, {Zhou}, {Zhou}, {Zhou}, {Zhou}, {Zhou}, {Zhou}, {Zhu},
  {Zhu}, {Zhu}, {Zhu}, \& {Zuo}}]{crab21}
{Cao}, Z., {Aharonian}, F., {An}, Q., {et~al.} 2021{\natexlab{a}}, Science,
  373, 425

\bibitem[{{Cao} {et~al.}(2021{\natexlab{b}}){Cao}, {Aharonian}, {An},
  {Axikegu}, {Bai}, {Bai}, {Bao}, {Bastieri}, {Bi}, {Bi}, {Cai}, {Cai}, {Cao},
  {Chang}, {Chang}, {Chen}, {Chen}, {Chen}, {Chen}, {Chen}, {Chen}, {Chen},
  {Chen}, {Chen}, {Chen}, {Chen}, {Chen}, {Chen}, {Chen}, {Cheng}, {Cheng},
  {Cui}, {Cui}, {Cui}, {D'Ettorre Piazzoli}, {Dai}, {Dai}, {Dai},
  {Danzengluobu}, {Volpe}, {Dong}, {Duan}, {Fan}, {Fan}, {Fan}, {Fang}, {Fang},
  {Feng}, {Feng}, {Feng}, {Feng}, {Gao}, {Gao}, {Gao}, {Gao}, {Gao}, {Ge},
  {Geng}, {Gong}, {Gou}, {Gu}, {Guo}, {Guo}, {Guo}, {Guo}, {Guo}, {Han}, {He},
  {He}, {He}, {He}, {He}, {He}, {Heller}, {Hor}, {Hou}, {Hu}, {Hu}, {Hu}, {Hu},
  {Huang}, {Huang}, {Huang}, {Huang}, {Huang}, {Huang}, {Ji}, {Ji}, {Jia},
  {Jiang}, {Jiang}, {Jin}, {Ke}, {Kuleshov}, {Levochkin}, {Li}, {Li}, {Li},
  {Li}, {Li}, {Li}, {Li}, {Li}, {Li}, {Li}, {Li}, {Li}, {Li}, {Li}, {Li}, {Li},
  {Li}, {Liang}, {Liang}, {Lin}, {Liu}, {Liu}, {Liu}, {Liu}, {Liu}, {Liu},
  {Liu}, {Liu}, {Liu}, {Liu}, {Liu}, {Liu}, {Liu}, {Liu}, {Liu}, {Liu}, {Long},
  {Lu}, {Lv}, {Ma}, {Ma}, {Ma}, {Mao}, {Masood}, {Min}, {Mitthumsiri},
  {Montaruli}, {Nan}, {Pang}, {Pattarakijwanich}, {Pei}, {Qi}, {Qi}, {Qiao},
  {Qin}, {Ruffolo}, {Rulev}, {S{\'a}iz}, {Shao}, {Shchegolev}, {Sheng}, {Shi},
  {Song}, {Stenkin}, {Stepanov}, {Su}, {Sun}, {Sun}, {Sun}, {Tam}, {Tang},
  {Tian}, {Wang}, {Wang}, {Wang}, {Wang}, {Wang}, {Wang}, {Wang}, {Wang},
  {Wang}, {Wang}, {Wang}, {Wang}, {Wang}, {Wang}, {Wang}, {Wang}, {Wang},
  {Wang}, {Wang}, {Wang}, {Wang}, {Wang}, {Wei}, {Wei}, {Wei}, {Wen}, {Wu},
  {Wu}, {Wu}, {Wu}, {Wu}, {Xi}, {Xia}, {Xia}, {Xiang}, {Xiao}, {Xiao}, {Xiao},
  {Xin}, {Xin}, {Xing}, {Xu}, {Xu}, {Xue}, {Yan}, {Yan}, {Yang}, {Yang},
  {Yang}, {Yang}, {Yang}, {Yang}, {Yang}, {Yao}, {Yao}, {Ye}, {Yin}, {Yin},
  {You}, {You}, {Yu}, {Yuan}, {Zeng}, {Zeng}, {Zeng}, {Zeng}, {Zha}, {Zhai},
  {Zhang}, {Zhang}, {Zhang}, {Zhang}, {Zhang}, {Zhang}, {Zhang}, {Zhang},
  {Zhang}, {Zhang}, {Zhang}, {Zhang}, {Zhang}, {Zhang}, {Zhang}, {Zhang},
  {Zhang}, {Zhang}, {Zhang}, {Zhao}, {Zhao}, {Zhao}, {Zhao}, {Zhao}, {Zheng},
  {Zheng}, {Zhou}, {Zhou}, {Zhou}, {Zhou}, {Zhou}, {Zhou}, {Zhu}, {Zhu}, {Zhu},
  {Zhu}, \& {Zuo}}]{cao21c}
{Cao}, Z., {Aharonian}, F., {An}, Q., {et~al.} 2021{\natexlab{b}}, \apjl, 917,
  L4

\bibitem[{{Cao} {et~al.}(2021{\natexlab{c}}){Cao}, {Aharonian}, {An},
  {Axikegu}, {Bai}, {Bao}, {Bastieri}, {Bi}, {Bi}, {Cai}, {Cai}, {Cao},
  {Chang}, {Chang}, {Chang}, {Chen}, {Chen}, {Chen}, {Chen}, {Chen}, {Chen},
  {Chen}, {Chen}, {Chen}, {Chen}, {Chen}, {Chen}, {Chen}, {Cheng}, {Cheng},
  {Cui}, {Cui}, {Cui}, {Dai}, {Dai}, {Dai}, {Danzengluobu}, {della Volpe},
  {D'Ettorre Piazzoli}, {Dong}, {Fan}, {Fan}, {Fan}, {Fang}, {Fang}, {Feng},
  {Feng}, {Feng}, {Feng}, {Gao}, {Gao}, {Gao}, {Gao}, {Ge}, {Geng}, {Gong},
  {Gou}, {Gu}, {Guo}, {Guo}, {Guo}, {Guo}, {Han}, {He}, {He}, {He}, {He}, {He},
  {He}, {Heller}, {Hor}, {Hou}, {Hou}, {Hu}, {Hu}, {Hu}, {Hu}, {Huang},
  {Huang}, {Huang}, {Huang}, {Huang}, {Ji}, {Ji}, {Jia}, {Jiang}, {Jiang},
  {Jin}, {Kuleshov}, {Levochkin}, {Li}, {Li}, {Li}, {Li}, {Li}, {Li}, {Li},
  {Li}, {Li}, {Li}, {Li}, {Li}, {Li}, {Li}, {Li}, {Li}, {Li}, {Liang}, {Liang},
  {Lin}, {Liu}, {Liu}, {Liu}, {Liu}, {Liu}, {Liu}, {Liu}, {Liu}, {Liu}, {Liu},
  {Liu}, {Liu}, {Liu}, {Liu}, {Liu}, {Long}, {Lu}, {Lv}, {Ma}, {Ma}, {Ma},
  {Mao}, {Masood}, {Mitthumsiri}, {Montaruli}, {Nan}, {Pang},
  {Pattarakijwanich}, {Pei}, {Qi}, {Ruffolo}, {Rulev}, {S{\'a}iz}, {Shao},
  {Shchegolev}, {Sheng}, {Shi}, {Song}, {Stenkin}, {Stepanov}, {Sun}, {Sun},
  {Sun}, {Tam}, {Tang}, {Tian}, {Wang}, {Wang}, {Wang}, {Wang}, {Wang}, {Wang},
  {Wang}, {Wang}, {Wang}, {Wang}, {Wang}, {Wang}, {Wang}, {Wang}, {Wang},
  {Wang}, {Wang}, {Wang}, {Wang}, {Wang}, {Wang}, {Wei}, {Wei}, {Wei}, {Wen},
  {Wu}, {Wu}, {Wu}, {Wu}, {Wu}, {Xi}, {Xia}, {Xia}, {Xiang}, {Xiao}, {Xiao},
  {Xin}, {Xin}, {Xing}, {Xu}, {Xu}, {Xue}, {Yan}, {Yang}, {Yang}, {Yang},
  {Yang}, {Yang}, {Yang}, {Yang}, {Yao}, {Yao}, {Ye}, {Yin}, {Yin}, {You},
  {You}, {Yu}, {Yuan}, {Zeng}, {Zeng}, {Zeng}, {Zeng}, {Zha}, {Zhai}, {Zhang},
  {Zhang}, {Zhang}, {Zhang}, {Zhang}, {Zhang}, {Zhang}, {Zhang}, {Zhang},
  {Zhang}, {Zhang}, {Zhang}, {Zhang}, {Zhang}, {Zhang}, {Zhang}, {Zhang},
  {Zhang}, {Zhang}, {Zhao}, {Zhao}, {Zhao}, {Zhao}, {Zhao}, {Zheng}, {Zheng},
  {Zhou}, {Zhou}, {Zhou}, {Zhou}, {Zhou}, {Zhou}, {Zhu}, {Zhu}, {Zhu}, {Zhu},
  \& {Zuo}}]{cao21a}
{Cao}, Z., {Aharonian}, F.~A., {An}, Q., {et~al.} 2021{\natexlab{c}}, \nat,
  594, 33

\bibitem[{{Cheng} \& {Zhang}(1996)}]{cheng96}
{Cheng}, K.~S. \& {Zhang}, J.~L. 1996, \apj, 463, 271

\bibitem[{{Cherenkov Telescope Array Consortium} {et~al.}(2019){Cherenkov
  Telescope Array Consortium}, {Acharya}, {Agudo}, {Al Samarai}, {Alfaro},
  {Alfaro}, {Alispach}, {Alves Batista}, {Amans}, {Amato}, {Ambrosi},
  {Antolini}, {Antonelli}, {Aramo}, {Araya}, {Armstrong}, {Arqueros},
  {Arrabito}, {Asano}, {Ashley}, {Backes}, {Balazs}, {Balbo}, {Ballester},
  {Ballet}, {Bamba}, {Barkov}, {Barres de Almeida}, {Barrio}, {Bastieri},
  {Becherini}, {Belfiore}, {Benbow}, {Berge}, {Bernardini}, {Bernardini},
  {Bernardos}, {Bernl{\"o}hr}, {Bertucci}, {Biasuzzi}, {Bigongiari}, {Biland},
  {Bissaldi}, {Biteau}, {Blanch}, {Blazek}, {Boisson}, {Bolmont}, {Bonanno},
  {Bonardi}, {Bonavolont{\`a}}, {Bonnoli}, {Bosnjak}, {B{\"o}ttcher},
  {Braiding}, {Bregeon}, {Brill}, {Brown}, {Brun}, {Brunetti}, {Buanes},
  {Buckley}, {Bugaev}, {B{\"u}hler}, {Bulgarelli}, {Bulik}, {Burton},
  {Burtovoi}, {Busetto}, {Canestrari}, {Capalbi}, {Capitanio}, {Caproni},
  {Caraveo}, {C{\'a}rdenas}, {Carlile}, {Carosi}, {Carqu{\'\i}n}, {Carr},
  {Casanova}, {Cascone}, {Catalani}, {Catalano}, {Cauz}, {Cerruti}, {Chadwick},
  {Chaty}, {Chaves}, {Chen}, {Chen}, {Chernyakova}, {Chikawa}, {Christov},
  {Chudoba}, {Cie{\'s}lar}, {Coco}, {Colafrancesco}, {Colin}, {Conforti},
  {Connaughton}, {Conrad}, {Contreras}, {Cortina}, {Costa}, {Costantini},
  {Cotter}, {Covino}, {Crocker}, {Cuadra}, {Cuevas}, {Cumani}, {D'A{\`\i}},
  {D'Ammando}, {D'Avanzo}, {D'Urso}, {Daniel}, {Davids}, {Dawson}, {Dazzi}, {De
  Angelis}, {de C{\'a}ssia dos Anjos}, {De Cesare}, {De Franco}, {de Gouveia
  Dal Pino}, {de la Calle}, {de los Reyes Lopez}, {De Lotto}, {De Luca}, {De
  Lucia}, {de Naurois}, {de O{\~n}a Wilhelmi}, {De Palma}, {De Persio}, {de
  Souza}, {Deil}, {Del Santo}, {Delgado}, {della Volpe}, {Di Girolamo}, {Di
  Pierro}, {Di Venere}, {D{\'\i}az}, {Dib}, {Diebold}, {Djannati-Ata{\"\i}},
  {Dom{\'\i}nguez}, {Dominis Prester}, {Dorner}, {Doro}, {Drass}, {Dravins},
  {Dubus}, {Dwarkadas}, {Ebr}, {Eckner}, {Egberts}, {Einecke}, {Ekoume},
  {Els{\"a}sser}, {Ernenwein}, {Espinoza}, {Evoli}, {Fairbairn},
  {Falceta-Goncalves}, {Falcone}, {Farnier}, {Fasola}, {Fedorova}, {Fegan},
  {Fernandez-Alonso}, {Fern{\'a}ndez-Barral}, {Ferrand}, {Fesquet},
  {Filipovic}, {Fioretti}, {Fontaine}, {Fornasa}, {Fortson}, {Freixas
  Coromina}, {Fruck}, {Fujita}, {Fukazawa}, {Funk}, {F{\"u}{\ss}ling},
  {Gabici}, {Gadola}, {Gallant}, {Garcia}, {Garcia L{\'o}pez}, {Garczarczyk},
  {Gaskins}, {Gasparetto}, {Gaug}, {Gerard}, {Giavitto}, {Giglietto}, {Giommi},
  {Giordano}, {Giro}, {Giroletti}, {Giuliani}, {Glicenstein}, {Gnatyk},
  {Godinovic}, {Goldoni}, {G{\'o}mez-Vargas}, {Gonz{\'a}lez}, {Gonz{\'a}lez},
  {G{\"o}tz}, {Graham}, {Grandi}, {Granot}, {Green}, {Greenshaw}, {Griffiths},
  {Gunji}, {Hadasch}, {Hara}, {Hardcastle}, {Hassan}, {Hayashi}, {Hayashida},
  {Heller}, {Helo}, {Hermann}, {Hinton}, {Hnatyk}, {Hofmann}, {Holder},
  {Horan}, {H{\"o}randel}, {Horns}, {Horvath}, {Hovatta}, {Hrabovsky},
  {Hrupec}, {Humensky}, {H{\"u}tten}, {Iarlori}, {Inada}, {Inome}, {Inoue},
  {Inoue}, {Inoue}, {Iocco}, {Ioka}, {Iori}, {Ishio}, {Iwamura}, {Jamrozy},
  {Janecek}, {Jankowsky}, {Jean}, {Jung-Richardt}, {Jurysek}, {Kaaret},
  {Karkar}, {Katagiri}, {Katz}, {Kawanaka}, {Kazanas}, {Kh{\'e}lifi}, {Kieda},
  {Kimeswenger}, {Kimura}, {Kisaka}, {Knapp}, {Kn{\"o}dlseder}, {Koch},
  {Kohri}, {Komin}, {Kosack}, {Kraus}, {Krause}, {Krau{\ss}}, {Kubo}, {Kukec
  Mezek}, {Kuroda}, {Kushida}, {La Palombara}, {Lamanna}, {Lang}, {Lapington},
  {Le Blanc}, {Leach}, {Lees}, {Lefaucheur}, {Leigui de Oliveira}, {Lenain},
  {Lico}, {Limon}, {Lindfors}, {Lohse}, {Lombardi}, {Longo}, {L{\'o}pez},
  {L{\'o}pez-Coto}, {Lu}, {Lucarelli}, {Luque-Escamilla}, {Lyard}, {Maccarone},
  {Maier}, {Majumdar}, {Malaguti}, {Mandat}, {Maneva}, {Manganaro}, {Mangano},
  {Marcowith}, {Mar{\'\i}n}, {Markoff}, {Mart{\'\i}}, {Martin},
  {Mart{\'\i}nez}, {Mart{\'\i}nez}, {Masetti}, {Masuda}, {Maurin}, {Maxted},
  {Mazin}, {Medina}, {Melandri}, {Mereghetti}, {Meyer}, {Minaya}, {Mirabal},
  {Mirzoyan}, {Mitchell}, {Mizuno}, {Moderski}, {Mohammed}, {Mohrmann},
  {Montaruli}, {Moralejo}, {Morcuende-Parrilla}, {Mori}, {Morlino}, {Morris},
  {Morselli}, {Moulin}, {Mukherjee}, {Mundell}, {Murach}, {Muraishi}, {Murase},
  {Nagai}, {Nagataki}, {Nagayoshi}, {Naito}, {Nakamori}, {Nakamura}, {Niemiec},
  {Nieto}, {Niko{\l}ajuk}, {Nishijima}, {Noda}, {Nosek}, {Novosyadlyj},
  {Nozaki}, {O'Brien}, {Oakes}, {Ohira}, {Ohishi}, {Ohm}, {Okazaki}, {Okumura},
  {Ong}, {Orienti}, {Orito}, {Osborne}, {Ostrowski}, {Otte}, {Oya}, {Padovani},
  {Paizis}, {Palatiello}, {Palatka}, {Paoletti}, {Paredes}, {Pareschi},
  {Parsons}, {Pe'er}, {Pech}, {Pedaletti}, {Perri}, {Persic}, {Petrashyk},
  {Petrucci}, {Petruk}, {Peyaud}, {Pfeifer}, {Piano}, {Pisarski}, {Pita},
  {Pohl}, {Polo}, {Pozo}, {Prandini}, {Prast}, {Principe}, {Prokhorov},
  {Prokoph}, {Prouza}, {P{\"u}hlhofer}, {Punch}, {P{\"u}rckhauer}, {Queiroz},
  {Quirrenbach}, {Rain{\`o}}, {Razzaque}, {Reimer}, {Reimer}, {Reisenegger},
  {Renaud}, {Rezaeian}, {Rhode}, {Ribeiro}, {Rib{\'o}}, {Richtler}, {Rico},
  {Rieger}, {Riquelme}, {Rivoire}, {Rizi}, {Rodriguez}, {Rodriguez Fernandez},
  {Rodr{\'\i}guez V{\'a}zquez}, {Rojas}, {Romano}, {Romeo}, {Rosado}, {Rovero},
  {Rowell}, {Rudak}, {Rugliancich}, {Rulten}, {Sadeh}, {Safi-Harb}, {Saito},
  {Sakaki}, {Sakurai}, {Salina}, {S{\'a}nchez-Conde}, {Sandaker}, {Sandoval},
  {Sangiorgi}, {Sanguillon}, {Sano}, {Santander}, {Sarkar}, {Satalecka},
  {Saturni}, {Schioppa}, {Schlenstedt}, {Schneider}, {Schoorlemmer},
  {Schovanek}, {Schulz}, {Schussler}, {Schwanke}, {Sciacca}, {Scuderi},
  {Seitenzahl}, {Semikoz}, {Sergijenko}, {Servillat}, {Shalchi}, {Shellard},
  {Sidoli}, {Siejkowski}, {Sillanp{\"a}{\"a}}, {Sironi}, {Sitarek}, {Sliusar},
  {Slowikowska}, {Sol}, {Stamerra}, {Stani{\v{c}}}, {Starling}, {Stawarz},
  {Stefanik}, {Stephan}, {Stolarczyk}, {Stratta}, {Straumann}, {Suomijarvi},
  {Supanitsky}, {Tagliaferri}, {Tajima}, {Tavani}, {Tavecchio}, {Tavernet},
  {Tayabaly}, {Tejedor}, {Temnikov}, {Terada}, {Terrier}, {Terzic}, {Teshima},
  {Testa}, {Thoudam}, {Tian}, {Tibaldo}, {Tluczykont}, {Todero Peixoto},
  {Tokanai}, {Tomastik}, {Tonev}, {Tornikoski}, {Torres}, {Torresi}, {Tosti},
  {Tothill}, {Tovmassian}, {Travnicek}, {Trichard}, {Trifoglio}, {Troyano
  Pujadas}, {Tsujimoto}, {Umana}, {Vagelli}, {Vagnetti}, {Valentino},
  {Vallania}, {Valore}, {van Eldik}, {Vandenbroucke}, {Varner}, {Vasileiadis},
  {Vassiliev}, {V{\'a}zquez Acosta}, {Vecchi}, {Vega}, {Vercellone}, {Veres},
  {Vergani}, {Verzi}, {Vettolani}, {Viana}, {Vigorito}, {Villanueva}, {Voelk},
  {Vollhardt}, {Vorobiov}, {Vrastil}, {Vuillaume}, {Wagner}, {Wagner},
  {Walter}, {Ward}, {Warren}, {Watson}, {Werner}, {White}, {White},
  {Wierzcholska}, {Wilcox}, {Will}, {Williams}, {Wischnewski}, {Wood},
  {Yamamoto}, {Yamazaki}, {Yanagita}, {Yang}, {Yoshida}, {Yoshiike},
  {Yoshikoshi}, {Zacharias}, {Zaharijas}, {Zampieri}, {Zandanel}, {Zanin},
  {Zavrtanik}, {Zavrtanik}, {Zdziarski}, {Zech}, {Zechlin}, {Zhdanov},
  {Ziegler}, \& {Zorn}}]{CTA19}
{Cherenkov Telescope Array Consortium}, {Acharya}, B.~S., {Agudo}, I., {et~al.}
  2019, {Science with the Cherenkov Telescope Array}

\bibitem[{{de O{\~n}a Wilhelmi} {et~al.}(2022){de O{\~n}a Wilhelmi},
  {L{\'o}pez-Coto}, {Amato}, \& {Aharonian}}]{emma22}
{de O{\~n}a Wilhelmi}, E., {L{\'o}pez-Coto}, R., {Amato}, E., \& {Aharonian},
  F. 2022, \apjl, 930, L2

\bibitem[{{De Sarkar}(2023)}]{desarkar23}
{De Sarkar}, A. 2023, \mnras, 521, L5

\bibitem[{{De Sarkar} {et~al.}(2021){De Sarkar}, {Biswas}, \&
  {Gupta}}]{desarkar21}
{De Sarkar}, A., {Biswas}, S., \& {Gupta}, N. 2021, Journal of High Energy
  Astrophysics, 29, 1

\bibitem[{{De Sarkar} \& {Gupta}(2022)}]{desarkar22a}
{De Sarkar}, A. \& {Gupta}, N. 2022, \apj, 934, 118

\bibitem[{{De Sarkar} {et~al.}(2022{\natexlab{a}}){De Sarkar}, {Roy},
  {Majumdar}, {Gupta}, {Brunthaler}, {Menten}, {Dzib}, {Medina}, \&
  {Wyrowski}}]{desarkar22c}
{De Sarkar}, A., {Roy}, N., {Majumdar}, P., {et~al.} 2022{\natexlab{a}}, \apjl,
  927, L35

\bibitem[{{De Sarkar} {et~al.}(2022{\natexlab{b}}){De Sarkar}, {Zhang},
  {Mart{\'\i}n}, {Torres}, {Li}, \& {Hou}}]{desarkar22b}
{De Sarkar}, A., {Zhang}, W., {Mart{\'\i}n}, J., {et~al.} 2022{\natexlab{b}},
  \aap, 668, A23

\bibitem[{Dogiel {et~al.}(2015)Dogiel, Chernyshov, Kiselev, Nobukawa, Cheng,
  Hui, Ko, Nobukawa, \& Tsuru}]{dogiel15}
Dogiel, V.~A., Chernyshov, D.~O., Kiselev, A.~M., {et~al.} 2015, \apj, 809, 48

\bibitem[{{Fang}(2022)}]{fang22}
{Fang}, K. 2022, Frontiers in Astronomy and Space Sciences, 9, 1022100

\bibitem[{{Fujita} {et~al.}(2009){Fujita}, {Ohira}, {Tanaka}, \&
  {Takahara}}]{fujita09}
{Fujita}, Y., {Ohira}, Y., {Tanaka}, S.~J., \& {Takahara}, F. 2009, \apjl, 707,
  L179

\bibitem[{{Gabici} {et~al.}(2009){Gabici}, {Aharonian}, \&
  {Casanova}}]{gabici09}
{Gabici}, S., {Aharonian}, F.~A., \& {Casanova}, S. 2009, \mnras, 396, 1629

\bibitem[{{Gaensler} \& {Slane}(2006)}]{gaensler06}
{Gaensler}, B.~M. \& {Slane}, P.~O. 2006, \araa, 44, 17

\bibitem[{{Ghisellini} {et~al.}(1988){Ghisellini}, {Guilbert}, \&
  {Svensson}}]{ghisellini}
{Ghisellini}, G., {Guilbert}, P.~W., \& {Svensson}, R. 1988, \apjl, 334, L5

\bibitem[{{Giacinti} {et~al.}(2020){Giacinti}, {Mitchell}, {L{\'o}pez-Coto},
  {Joshi}, {Parsons}, \& {Hinton}}]{giacinti20}
{Giacinti}, G., {Mitchell}, A.~M.~W., {L{\'o}pez-Coto}, R., {et~al.} 2020,
  \aap, 636, A113

\bibitem[{{Goldreich} \& {Julian}(1969)}]{goldreich69}
{Goldreich}, P. \& {Julian}, W.~H. 1969, \apj, 157, 869

\bibitem[{Hahn(2016)}]{hahn}
Hahn, J. 2016, PoS, ICRC2015, 917

\bibitem[{{Kafexhiu} {et~al.}(2014){Kafexhiu}, {Aharonian}, {Taylor}, \&
  {Vila}}]{kafe}
{Kafexhiu}, E., {Aharonian}, F., {Taylor}, A.~M., \& {Vila}, G.~S. 2014, \prd,
  90, 123014

\bibitem[{{Kar} \& {Gupta}(2022)}]{kar22}
{Kar}, A. \& {Gupta}, N. 2022, \apj, 926, 110

\bibitem[{{Kelner} {et~al.}(2015){Kelner}, {Prosekin}, \&
  {Aharonian}}]{kelner15}
{Kelner}, S.~R., {Prosekin}, A.~Y., \& {Aharonian}, F.~A. 2015, \aj, 149, 33

\bibitem[{{Li} {et~al.}(2021){Li}, {Liu}, {de O{\~n}a Wilhelmi}, {Torres},
  {Liu}, {Kerr}, {B{\"u}hler}, {Su}, {He}, \& {Xiao}}]{li21}
{Li}, J., {Liu}, R.-Y., {de O{\~n}a Wilhelmi}, E., {et~al.} 2021, The
  Astrophysical Journal: Letters, 913, L33

\bibitem[{{Liu}(2022)}]{ruo22}
{Liu}, R.-Y. 2022, International Journal of Modern Physics A, 37, 2230011

\bibitem[{{Liu} {et~al.}(2019{\natexlab{a}}){Liu}, {Ge}, {Sun}, \&
  {Wang}}]{ruo19b}
{Liu}, R.-Y., {Ge}, C., {Sun}, X.-N., \& {Wang}, X.-Y. 2019{\natexlab{a}},
  \apj, 875, 149

\bibitem[{{Liu} {et~al.}(2019{\natexlab{b}}){Liu}, {Yan}, \& {Zhang}}]{ruo19}
{Liu}, R.-Y., {Yan}, H., \& {Zhang}, H. 2019{\natexlab{b}}, \prl, 123, 221103

\bibitem[{{Lyutikov} \& {Gavriil}(2006)}]{lyutikov06}
{Lyutikov}, M. \& {Gavriil}, F.~P. 2006, \mnras, 368, 690

\bibitem[{{Makino} {et~al.}(2019){Makino}, {Fujita}, {Nobukawa}, {Matsumoto},
  \& {Ohira}}]{makino19}
{Makino}, K., {Fujita}, Y., {Nobukawa}, K.~K., {Matsumoto}, H., \& {Ohira}, Y.
  2019, \pasj, 71, 78

\bibitem[{{Ohira} {et~al.}(2010){Ohira}, {Murase}, \& {Yamazaki}}]{ohira10}
{Ohira}, Y., {Murase}, K., \& {Yamazaki}, R. 2010, \aap, 513, A17

\bibitem[{Ohira {et~al.}(2012)Ohira, Yamazaki, Kawanaka, \& Ioka}]{ohira12}
Ohira, Y., Yamazaki, R., Kawanaka, N., \& Ioka, K. 2012, \mnras, 427, 91

\bibitem[{Popescu {et~al.}(2017)Popescu, Yang, Tuffs, Natale, Rushton, \&
  Aharonian}]{popescu}
Popescu, C.~C., Yang, R., Tuffs, R.~J., {et~al.} 2017, \mnras, 470, 2539

\bibitem[{{Prosekin} {et~al.}(2015){Prosekin}, {Kelner}, \&
  {Aharonian}}]{prosekin15}
{Prosekin}, A.~Y., {Kelner}, S.~R., \& {Aharonian}, F.~A. 2015, \prd, 92,
  083003

\bibitem[{{Smith} {et~al.}(2023){Smith}, {Bruel}, {Clark}, {Guillemot}, {Kerr},
  {Ray}, {Abdollahi}, {Ajello}, {Baldini}, {Ballet}, {Baring}, {Bassa},
  {Becerra Gonzalez}, {Bellazzini}, {Berretta}, {Bhattacharyya}, {Bissaldi},
  {Bonino}, {Bottacini}, {Bregeon}, {Burgay}, {Burnett}, {Cameron}, {Camilo},
  {Caputo}, {Caraveo}, {Cavazzuti}, {Chiaro}, {Ciprini}, {Cognard},
  {Cristarella Orestano}, {Crnogorcevic}, {Cuoco}, {Cutini}, {D'Ammando}, {de
  Angelis}, {De Gaetano}, {de Menezes}, {de Palma}, {DeCesar}, {Deneva}, {Di
  Lalla}, {Di Venere}, {Fana Dirirsa}, {Dominguez}, {Dumora}, {Fegan},
  {Ferrara}, {Fiori}, {Fleischhack}, {Flynn}, {Franckowiak}, {Freire},
  {Fukazawa}, {Fusco}, {Galanti}, {Gammaldi}, {Gargano}, {Gasparrini},
  {Giacchino}, {Giglietto}, {Giordano}, {Giroletti}, {Green}, {Grenier},
  {Guiriec}, {Gustafsson}, {Harding}, {Hays}, {Hewitt}, {Horan}, {Hou},
  {Jankowski}, {Johnson}, {Johnson}, {Johnston}, {Kataoka}, {Keith}, {Kramer},
  {Kuss}, {Latronico}, {Shiu-Hang}, {Lee}, {Li}, {Li}, {Limyansky}, {Longo},
  {Loparco}, {Lorusso}, {Lovellette}, {Lower}, {Lubrano}, {Lyne}, {Maldera},
  {Manchester}, {Manfreda}, {Marelli}, {Marta-Devesa}, {Mazziotta}, {McEnery},
  {Mereu}, {Michelson}, {Mitthumsiri}, {Mizuno}, {Moiseev}, {Monzani},
  {Morselli}, {Negro}, {Nemmen}, {Nieder}, {Nuss}, {Omodei}, {Orienti},
  {Orlando}, {Ormes}, {Palatiello}, {Paneque}, {Panzarini}, {Persic},
  {Pesce-Rollins}, {Pillera}, {Poon}, {Porter}, {Principe}, {Raino}, {Rando},
  {Ransom}, {Razzano}, {Razzaque}, {Reimer}, {Reimer}, {Renault-Tinacci},
  {Romani}, {Sanchez-Conde}, {Saz Parkinson}, {Scotton}, {Serini}, {Sgro},
  {Shannon}, {Sharma}, {Siskind}, {Spandre}, {Spinelli}, {Stappers},
  {Stephens}, {Suson}, {Tajima}, {Tak}, {Theureau}, {Thompson}, {Tibolla},
  {Torres}, {Valverde}, {Venter}, {Wadiasingh}, {Wang}, {Wang}, {Weltevrede},
  {Wood}, \& {Zaharijas}}]{smith23}
{Smith}, D.~A., {Bruel}, P., {Clark}, C.~J., {et~al.} 2023, arXiv e-prints,
  arXiv:2307.11132

\bibitem[{Su {et~al.}(2019)Su, Yang, Zhang, Gong, Wang, Zhou, Wang, Chen, Sun,
  Chen, Xu, \& Jiang}]{Su19}
Su, Y., Yang, J., Zhang, S., {et~al.} 2019, The Astrophysical Journal
  Supplement Series, 240, 9

\bibitem[{{Sudoh} {et~al.}(2019){Sudoh}, {Linden}, \& {Beacom}}]{sudoh19}
{Sudoh}, T., {Linden}, T., \& {Beacom}, J.~F. 2019, \prd, 100, 043016

\bibitem[{{Taylor} {et~al.}(1993){Taylor}, {Manchester}, \& {Lyne}}]{taylor93}
{Taylor}, J.~H., {Manchester}, R.~N., \& {Lyne}, A.~G. 1993, \apjs, 88, 529

\bibitem[{{Vercellone}(2023)}]{astri23}
{Vercellone}, S. 2023, arXiv e-prints, arXiv:2302.10000

\bibitem[{{Vigan{\`o}} \& {Torres}(2015)}]{vigano15b}
{Vigan{\`o}}, D. \& {Torres}, D.~F. 2015, \mnras, 449, 3755

\bibitem[{{Vigan{\`o}} {et~al.}(2015{\natexlab{a}}){Vigan{\`o}}, {Torres},
  {Hirotani}, \& {Pessah}}]{vigano15c}
{Vigan{\`o}}, D., {Torres}, D.~F., {Hirotani}, K., \& {Pessah}, M.~E.
  2015{\natexlab{a}}, \mnras, 447, 2631

\bibitem[{{Vigan{\`o}} {et~al.}(2015{\natexlab{b}}){Vigan{\`o}}, {Torres},
  {Hirotani}, \& {Pessah}}]{vigano15}
{Vigan{\`o}}, D., {Torres}, D.~F., {Hirotani}, K., \& {Pessah}, M.~E.
  2015{\natexlab{b}}, \mnras, 447, 1164

\bibitem[{{Wood} {et~al.}(2017){Wood}, {Caputo}, {Charles}, {Di Mauro},
  {Magill}, {Perkins}, \& {Fermi-LAT Collaboration}}]{wood17}
{Wood}, M., {Caputo}, R., {Charles}, E., {et~al.} 2017, in International Cosmic
  Ray Conference, Vol. 301, 35th International Cosmic Ray Conference
  (ICRC2017), 824

\bibitem[{{Xu} {et~al.}(2016){Xu}, {Yan}, \& {Lazarian}}]{xu16}
{Xu}, S., {Yan}, H., \& {Lazarian}, A. 2016, \apj, 826, 166

\end{thebibliography}
%

\end{document}